\documentclass[aps,prl,reprint,superscriptaddress]{revtex4-2}

\usepackage[normalem]{ulem}
\usepackage{graphicx}
\usepackage{dcolumn}
\usepackage{bm}
\usepackage{color}
\usepackage{comment}
\usepackage{amsmath}
\usepackage{amssymb}
\usepackage{hyperref}

\begin{document}

\title{Giant nonlinear Hall effect in a Pt/ferrimagnetic insulator bilayer under Zeeman--exchange frustration}

\author{Takayuki~Shiino}
\thanks{These authors contributed equally to this work.}
\email{tshiino@icmab.es}
\affiliation{Institut de Ciència de Materials de Barcelona (ICMAB-CSIC), Carrer dels Til.lers, 08193 Cerdanyola del Vallès, Spain}

\author{Matteo~Fettizio}
\thanks{These authors contributed equally to this work.}
\affiliation{Institut de Ciència de Materials de Barcelona (ICMAB-CSIC), Carrer dels Til.lers, 08193 Cerdanyola del Vallès, Spain}

\author{Weronika~Janus}
\affiliation{Institut de Ciència de Materials de Barcelona (ICMAB-CSIC), Carrer dels Til.lers, 08193 Cerdanyola del Vallès, Spain}

\author{Can~Onur~Avci}
\email{cavci@icmab.es}
\affiliation{Institut de Ciència de Materials de Barcelona (ICMAB-CSIC), Carrer dels Til.lers, 08193 Cerdanyola del Vallès, Spain}

\date{\today}

\begin{abstract}
Competing magnetic interactions can create metastable or unstable states and render magnetic systems highly susceptible to external perturbations. Here we show that Zeeman–exchange frustration in an Al-substituted terbium iron garnet with a compositional gradient across its thickness gives rise to a giant nonlinear Hall response in an adjacent Pt layer. Near magnetic compensation, a field-induced spin-flip transition is accompanied by unusually large higher-order odd harmonic voltages, with the third, fifth, and seventh harmonics reaching amplitudes comparable to that of the first harmonic. The field, temperature, and current dependences collectively identify Joule heating as the parametric drive of the harmonic response. Macrospin-chain simulations further show that current-induced thermal modulation periodically switches the interfacial Fe magnetization between exchange- and Zeeman-dominated states and reproduces the observed harmonic signals. These results demonstrate how frustration can convert a weak thermal perturbation into a large nonlinear electrical response, providing a route to nonlinear magnetotransport in compensated ferrimagnets.
\end{abstract}

\maketitle

Magnetic states in solids are governed by competing energy terms, including exchange, anisotropy, and Zeeman energies, whose delicate balance can give rise to metastability and highly nonlinear dynamics. Driven nonlinear magnetic phenomena have been widely explored in magnonics \cite{Zheng2023tutorial, Demokritov2006bose}, spin-torque auto-oscillation \cite{Kiselev2003microwave, Demidov2012magnetic}, and dynamical stabilization of magnetic states \cite{Lerose2019prethermal, Kulikov2022kapitza, Kurebayashi2026dynamical}. 
In harmonic magnetotransport, however, current-induced nonlinear signals are usually perturbative: an ac current weakly distorts the magnetization or resistance, producing higher harmonics much smaller than the fundamental \cite{garello2013symmetry,Avci2015unidirectional,he2018bilinear,Cheng2022third}. 
A different regime can arise when the current parametrically modulates an internal magnetic energy scale across a stability boundary. The response is then governed not by oscillations around a single equilibrium state, but by repeated loss of stability and switching between competing magnetic configurations. Such dynamics can yield a large nonlinear transport response that remains largely unexplored.

Compensated ferrimagnets provide a natural platform for competing magnetic energies and metastable states. In these materials, antiferromagnetically coupled sublattices of different elements can give rise to magnetic compensation and noncollinear spin states \cite{Neel1972, Dionne2009, Lahoubi2003double, Li2024giant}. This sublattice structure and compensation physics underlie several remarkable phenomena, including ultrafast all-optical switching \cite{Kirilyuk2010ultrafast}, fast domain-wall motion \cite{Kim2017,Caretta2018}, and ultrastrong magnon--magnon coupling \cite{Liensberger2019exchange}. A particularly intriguing situation arises when the compensation condition varies spatially: exchange-coupled regions can then favor different magnetic configurations under an applied field \cite{Deb2021tunable, Wang2021probing}. The resulting incompatibility between Zeeman and exchange preferences creates a frustrated magnetic state that can host field-driven instabilities and metastable configurations.

Here we investigate the transport consequences of Zeeman–exchange competition in vertically inhomogeneous Al-substituted terbium iron garnet/platinum (Al:TbIG/Pt) heterostructures with perpendicular magnetic anisotropy. The compositional gradient across Al:TbIG produces exchange-coupled Fe- and Tb-dominated regions that favor competing magnetic configurations under an applied field, leading to a spin-flip instability near magnetic compensation. Harmonic Hall measurements in the Pt overlayer reveal unusually large odd-order voltage harmonics near this instability, with amplitudes comparable to the fundamental. By combining the field, temperature, and current dependences with macrospin-chain simulations, we show that Joule-heating-induced thermal modulation periodically drives the interfacial magnetization switching between competing magnetic states, giving rise to a giant nonlinear Hall response.
These results reveal a mechanism by which Zeeman--exchange frustration, enabled by a spatial distribution of magnetic compensation, is converted into giant nonlinear magnetotransport through Joule-heating-induced parametric switching.

\textit{Sample and measurements.---}
The Al:TbIG/Pt heterostructures were from the sample series reported in Ref.~\cite{Shiino2025tunable}. Al:TbIG was grown on GGG(111) by high-temperature co-sputtering of stoichiometric TbIG and Al$_2$O$_3$. On top, a 2 nm-thick TbIG layer was deposited at high temperature, followed by \textit{in situ} 4 nm of Pt deposition at room temperature. The stack was patterned into Hall bars by optical lithography and ion-beam etching. As shown later, the Al concentration varies across the film thickness, leading to depth-dependent magnetic properties and compensation conditions. Harmonic Hall measurements were performed by applying an a.c. current through the Pt overlayer and detecting harmonic Hall voltages \(V_H^{n\omega}\) with a focus on the odd harmonics \(n=1,3,5,7\). The first harmonic Hall response originates from the anomalous Hall component of the transverse spin Hall magnetoresistance (SMR) \cite{Nakayama2013,avci2017NM} and is primarily sensitive to the Fe-sublattice magnetization near the Al:TbIG/Pt interface. Unless otherwise noted, measurements were carried out at \(f=337\) Hz in the 295 - 315 K temperature range. Further details on growth, structural characterization, device fabrication, and transport measurements are provided in Ref.~\cite{Shiino2025tunable} and the Supplemental Material (SM) \cite{SM}.

\begin{figure}[t]
\includegraphics[width=\linewidth]{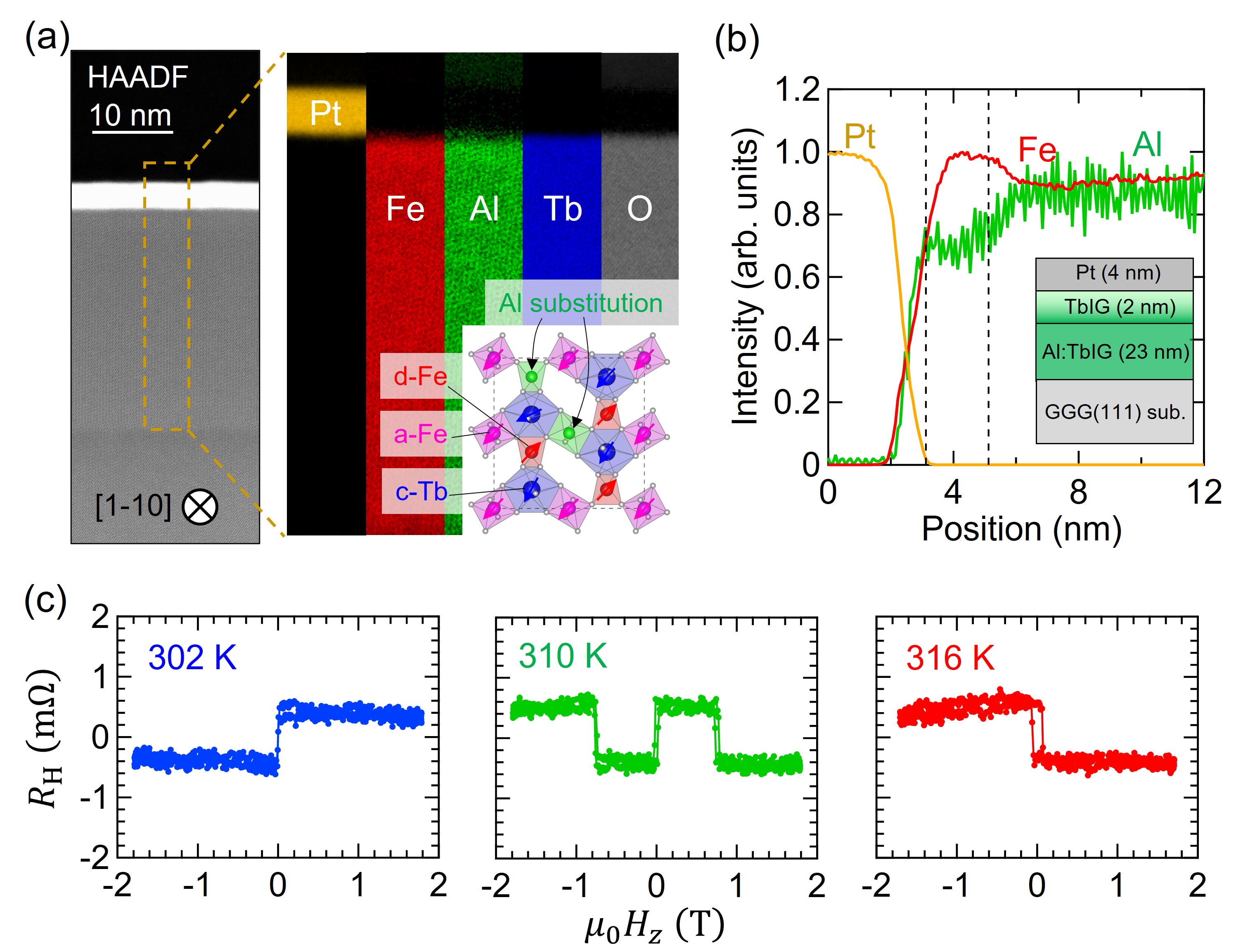}
\caption{
Structural characterization of the Al:TbIG/Pt heterostructure.
(a) Cross-sectional HAADF-STEM image and EELS elemental maps of Pt, Fe, Al, Tb, and O. 
Inset: crystal structure of Al-substituted TbIG, where Al$^{3+}$ partially replaces Fe$^{3+}$ at the $a$ and $d$ sites, drawn using VESTA \cite{Momma2011vesta}. 
(b) EELS depth profiles across the Pt/garnet interface. 
Vertical dashed lines mark the nominal 2-nm TbIG termination layer; Al remains finite in this region but is reduced relative to the underlying Al:TbIG layer. 
Inset: sputtering stack of the sample. 
The HAADF-STEM and EELS data are replotted from Ref.~\cite{Shiino2025tunable}.
(c) Hall resistance curves measured with low current at 302, 310 and 316 K. The ordinary Hall component has been subtracted. 
}
\label{FIG:1}
\end{figure}

Figure~\ref{FIG:1}(a,b) characterizes the vertical composition profile of the Al:TbIG/Pt heterostructure. The High-Angle Annular Dark-Field Scanning Transmission Electron Microscopy (HAADF-STEM) image and Electron Energy Loss Spectroscopy (EELS) analysis reveal that the Al concentration decreases toward the Pt interface rather than terminating abruptly at the additional 2 nm-thick TbIG layer. Since Al substitution modifies the balance between the Fe and Tb sublattices, this compositional gradient gives rise to depth-dependent magnetic properties, one of the most prominent consequences of the inhomogeneous composition. We therefore describe two distinct regions: an Al-poor interfacial region (probed by SMR in Pt) and an Al-rich bulk region underneath.

\begin{figure}[t]
\includegraphics[width=\linewidth]{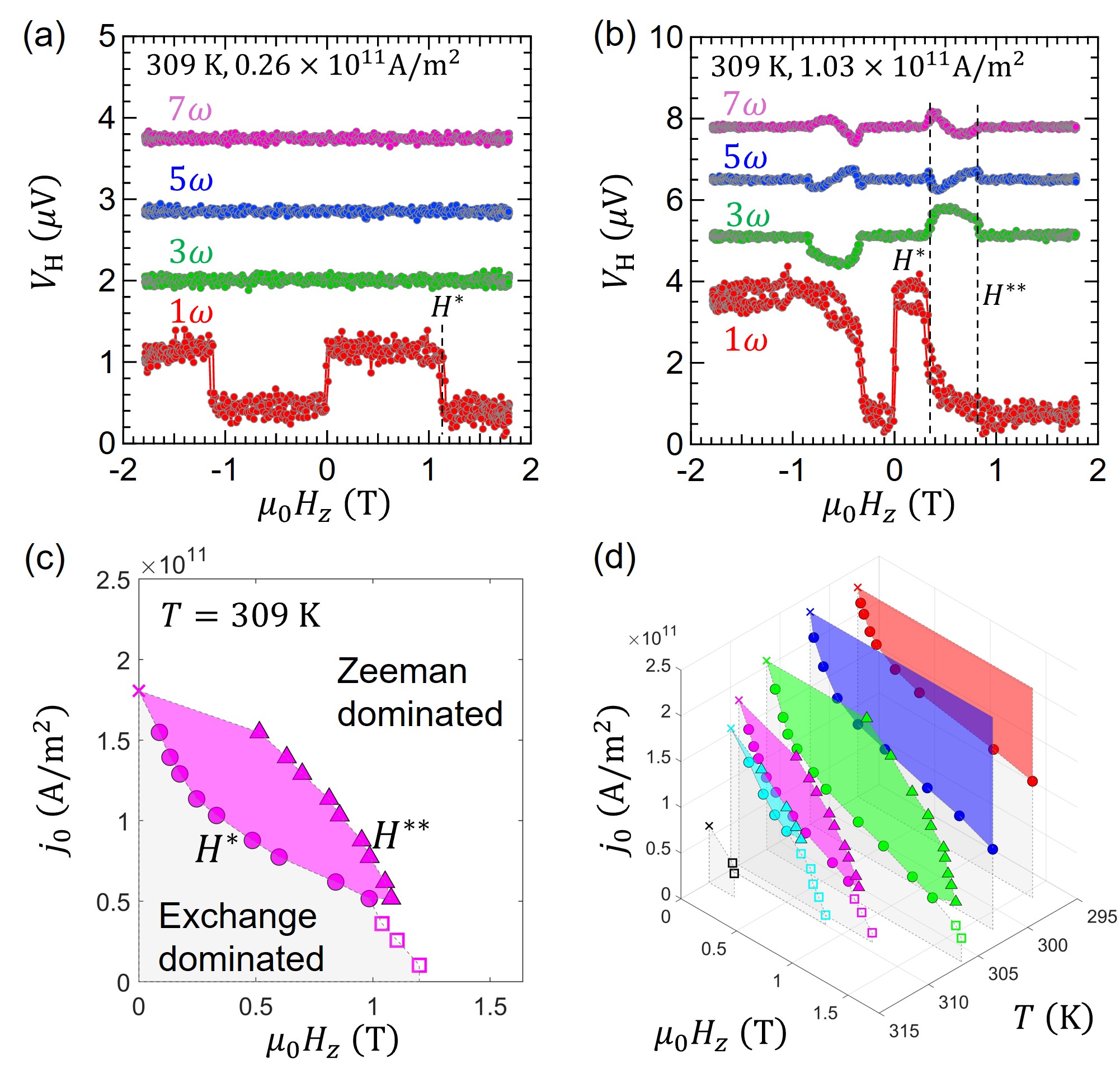}
\caption{
(a), (b) Real-part odd harmonic Hall voltages, \(V_H^{n\omega}\) (\(n=1,3,5,7\)), measured as a function of out-of-plane field at \(T=309\) K and the indicated current densities.
Dashed lines indicate \(H^\ast\) and, where observed, \(H^{\ast\ast}\).
Curves are vertically offset for clarity; the ordinary Hall contribution is subtracted from the \(1\omega\) data.
(c) \(j_0\)--\(H_z\) phase diagram at \(T=309\) K. 
Filled circles and triangles denote \(H^\ast\) and \(H^{\ast\ast}\), respectively; the shaded area indicates the nonlinear region. 
Open squares mark spin-flip-like transitions without nonlinear signals, and the cross indicates the absence of a resolved \(H^\ast\) transition.
(d) \(j_0\)--\(H_z\)--\(T\) phase diagram, with symbols as in (c). 
}
\label{FIG:2}
\end{figure}

Figure~\ref{FIG:1}(c) shows the first harmonic Hall resistances measured at $T=$ 302, 310, and 316 K. The sign reversal of the low-field Hall signal between 302 and 316 K indicates that the magnetization sensed by SMR at the Pt interface changes from Tb-dominated to Fe-dominated, suggesting that the effective magnetic compensation temperature lies between these two temperatures.\cite{Deb2021tunable,Wang2021probing,Shiino2025tunable,song2024temperature} While the 302 K and 316 K measurements exhibit the expected magnetic-field dependence, the intermediate-temperature curve at 310 K displays an unusual behavior. In addition to the low-field reversal, two additional full switching events appear at $\mu_0H\sim$0.8~T. The Hall signal therefore first switches at low field, then re-enters the opposite state at higher field, resembling a spin-flip behavior. The signal amplitude is inconsistent with a spin-flop transition (expected to set the signal at an intermediate state) and thus cannot be readily explained by a magnetically homogeneous ferrimagnet. Instead, it points to an unusual behavior that may be the consequence of distinct magnetic regions with different compensation conditions. Motivated by this observation and the structural characterization presented above, we consider a scenario in which the interfacial region probed by SMR has already become Fe-dominated at 310 K, while the underlying Al-rich region remains Tb-dominated. Exchange coupling forces the interfacial Fe-dominated region to follow the magnetization direction of the Tb-dominated bulk. An external magnetic field then introduces a competition between exchange and Zeeman energies because the field cannot simultaneously minimize the Zeeman energy of both regions. Above a threshold field, the interfacial magnetization flips into the field-favored configuration, producing the high-field switching observed in Fig.~\ref{FIG:1}(c). This field-induced instability constitutes the Zeeman–exchange competition that underpins the nonlinear phenomena discussed below.

Figure~\ref{FIG:2} characterizes the harmonic Hall response in the vicinity of the spin-flip signal. At \(T=309\) K and low current density, \(j_0=0.26\times10^{11}\) A/m\(^2\) [Fig.~\ref{FIG:2}(a)], only the first-harmonic exhibits the spin-flip transition, while the higher harmonics are absent. This indicates that the instability alone is insufficient to generate a sizable nonlinear response. Upon increasing the current density to \(j_0=1.03\times10^{11}\) A/m\(^2\) [Fig.~\ref{FIG:2}(b)], the response changes dramatically: large odd harmonic voltages emerge in a finite field interval beginning at \(H^\ast\) and vanishing above \(H^{\ast\ast}\). The confinement of the nonlinear response to the interval between $H^*$ and $H^{**}$ is a central experimental signature of the frustrated state. The \(3\omega\), \(5\omega\), and \(7\omega\) components attain amplitudes comparable to the change in the first-harmonic signal across the transition, far exceeding the magnitude expected from conventional harmonic Hall responses arising from current-induced variations of magnetic parameters such as the magnetization or anisotropy \cite{garello2013symmetry,Cheng2022third,Feng2026}. Higher odd harmonics beyond \(7\omega\) are also observed (see the SM \cite{SM}).

\begin{figure}[t]
\includegraphics[width=\linewidth]{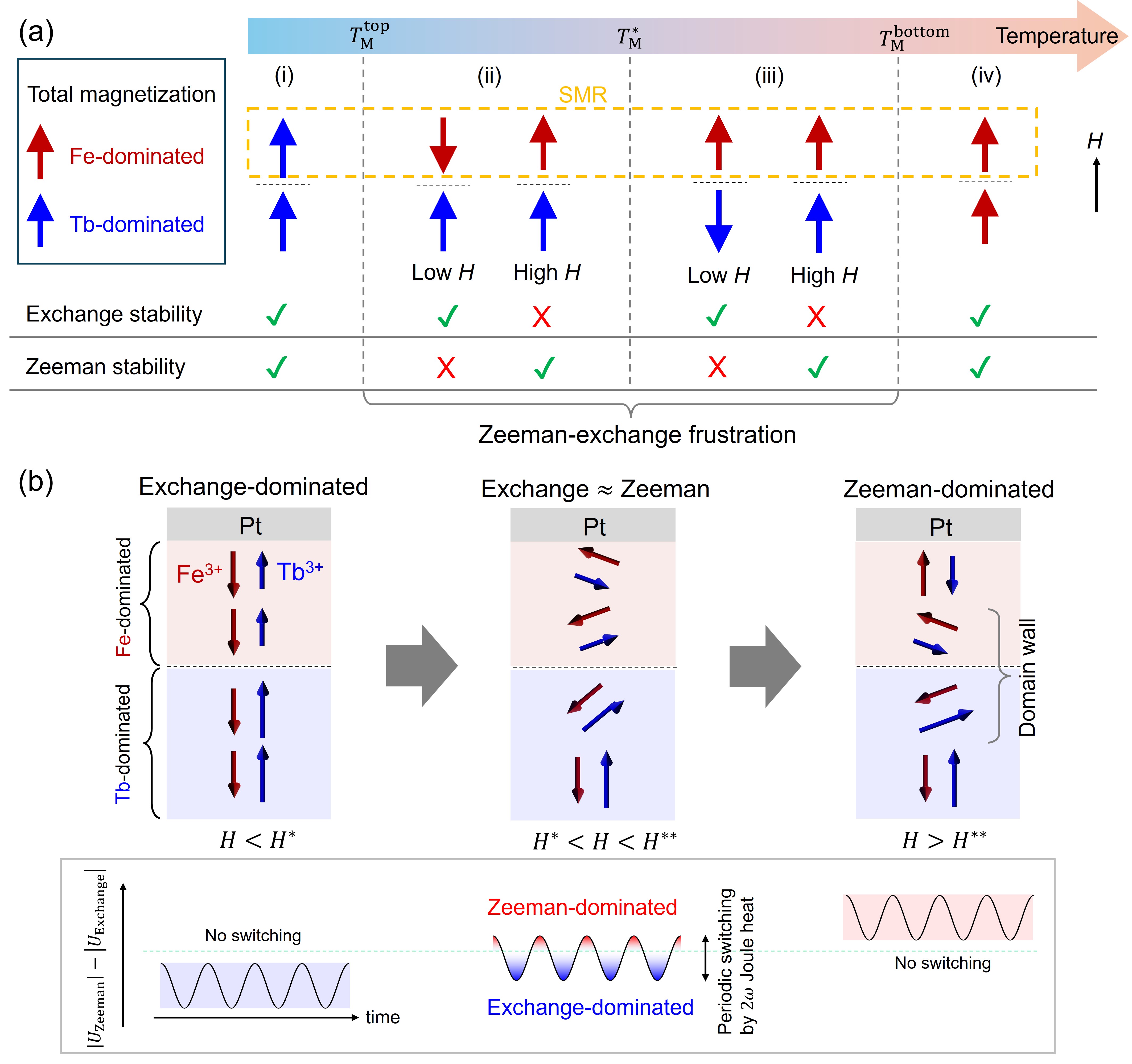}
\caption{
(a) Schematic magnetic configurations of the present vertically inhomogeneous ferrimagnet with \(T_{\rm M}^{\rm top}<T_{\rm M}^{\rm bottom}\). 
Red and blue arrows denote the net magnetization of Fe-dominated and Tb-dominated regions, respectively. 
The dashed yellow box indicates the interfacial region probed by SMR. 
Check marks and crosses indicate whether the exchange and Zeeman energies are stabilized for each configuration. 
This study focuses on region (ii), where exchange and Zeeman preferences compete and produce Zeeman--exchange frustration.
(b) Schematic spin configurations in region (ii), showing the exchange-dominated (\(H < H^\ast\)), boundary (\(H^\ast < H < H^{\ast\ast}\)), and Zeeman-dominated (\(H > H^{\ast\ast}\)) regimes. 
The bottom panel schematically illustrates the difference between the magnitudes of the Zeeman and exchange energies, $|U_{\rm Zeeman}| - |U_{\rm Exchange}|$. The energy (difference) oscillates due to oscillatory temperature by Joule heating. The green dashed line indicates the boundary separating the exchange- and Zeeman-dominated regions.
}
\label{FIG:3}
\end{figure}

\begin{figure}[t]
\includegraphics[width=\linewidth]{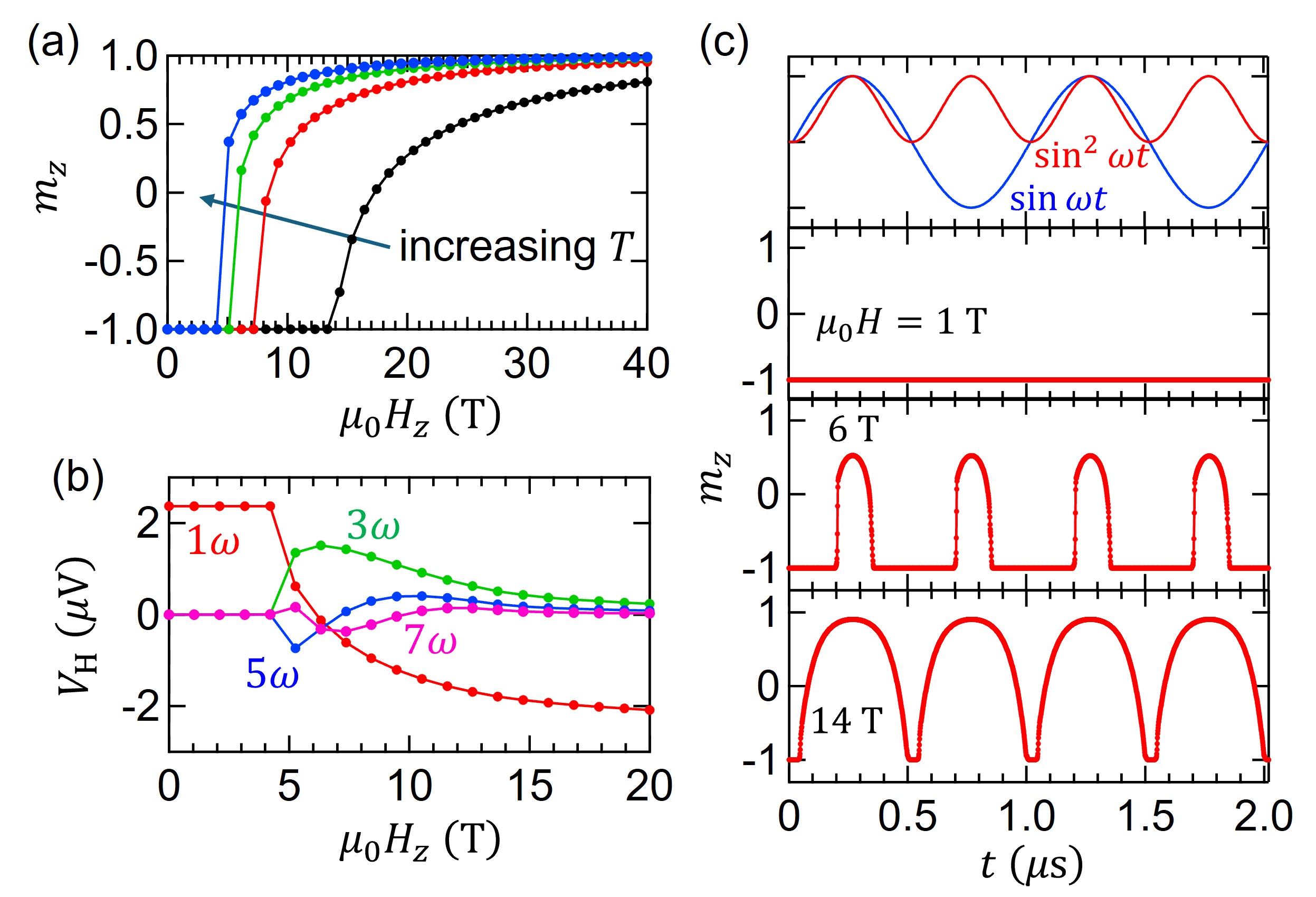}
\caption{
Numerical simulation of parametric switching.
(a) Static \(m_z\) of the interfacial macrospin as a function of \(\mu_0 H_z\) for increasing temperatures below \(T_M^\ast\), showing the reduction of the spin-flip field with temperature. 
(b) Simulated real-part odd harmonic Hall voltages \(V_H^{n\omega}\) (\(n=1,3,5,7\)) under Joule-heating-induced parametric modulation. 
(c) Time evolution of \(m_z\) at representative fields; the top panel shows \(\sin\omega t\) and \(\sin^2\omega t\). 
Periodic threshold crossing in the Zeeman--exchange-frustrated regime produces anharmonic switching and large odd harmonic voltages. 
}
\label{FIG:4}
\end{figure}

The variation of the harmonic response with current density is shown in Fig.~\ref{FIG:2}(c), while the same data taken at different temperatures are summarized in Fig.~\ref{FIG:2}(d). The evolution of the window between \(H^\ast\) and \(H^{\ast\ast}\) suggests an important role of Joule heating. A.c. Joule heating produces both a finite increase in the average sample temperature and an oscillatory temperature modulation. Estimates of the average temperature rise and additional Hall measurement results obtained at different temperatures and current densities are presented in the SM \cite{SM}. At temperatures well below compensation, relatively large currents are required to raise the average temperature into the Zeeman–exchange-frustrated regime where the spin-flip instability becomes accessible. At the same time, the larger current also increases the amplitude of the temperature modulation. As a result, the system explores a broader range of magnetic states during each current cycle, and progressively larger fields are required to fully stabilize the Zeeman-dominated configuration. Consequently, the field interval between $H^* - H^{**}$ expands with increasing current. The growth of this window is nevertheless limited by the accompanying d.c. temperature rise induced by Joule heating. As the average temperature approaches the effective compensation temperature $T_M^*$, the spin-flip instability shifts to lower fields and the Zeeman–exchange-frustrated regime shrinks. The resulting competition between the a.c. thermal modulation and the d.c. temperature increase governs the extent of the nonlinear region.

Figure~\ref{FIG:3}(a) summarizes the magnetic configurations expected in a vertically inhomogeneous Al:TbIG film. At low temperatures [region (i)], both the interfacial and bulk regions are Tb dominated, whereas at high temperatures [region (iv)] both are Fe dominated. Between these limits, the interfacial region crosses compensation first owing to its lower Al concentration. This gives rise to two intermediate regimes [regions (ii) and (iii)] in which the interface is Fe dominated while the bulk remains Tb dominated. The distinction between the two regimes lies in whether the net magnetization of the film is governed by the bulk [region (ii)] or by the interface [region (iii)]. Since SMR probes primarily the interfacial magnetization, the most pronounced anomalies are expected in region (ii), where exchange coupling forces the Fe-dominated interface to follow the Tb-dominated bulk. An external field then introduces a direct competition between Zeeman and exchange energies, leading to the spin-flip anomalies observed in Figs.~\ref{FIG:1} and~\ref{FIG:2}.

Figure~\ref{FIG:3}(b) illustrates the corresponding spin configurations across the film thickness in region (ii). For $H<H^*$, the magnetic state is exchange dominated. Between $H^*$ and $H^{**}$, Zeeman and exchange energies compete, producing a frustrated nonuniform spin texture. Above $H^{**}$, the interfacial region becomes fully Zeeman dominated and aligns with the applied field. These field-dependent spin configurations form the basis of the macrospin-chain model introduced below.

To capture the thickness-dependent magnetic configurations shown in Fig.~\ref{FIG:3}(b), we performed one-dimensional macrospin-chain simulations. The Al:TbIG film is represented by a chain of exchange-coupled spins with depth-dependent magnetic properties, reflecting the lower compensation temperature of the Al-poor interfacial region relative to the Al-rich bulk. The model incorporates Zeeman, perpendicular anisotropy, and nearest-neighbor exchange interactions using experimentally motivated temperature-dependent magnetic parameters and compensation temperatures. The Hall response is calculated from the topmost macrospin, corresponding to the interfacial Fe magnetization probed experimentally by SMR. To simulate current-induced effects, Joule heating is introduced through both a static temperature increase and a \(2\omega\) temperature modulation proportional to \(I^2(t)\), while spin-orbit torque is included separately to assess its contribution. Further details of the model are provided in the SM \cite{SM}.

Figure~\ref{FIG:4}(a) shows the calculated out-of-plane magnetization \(m_z\) of the topmost (interfacial) spin in the static limit. As the temperature approaches \(T_M^\ast\), the spin-flip transition shifts to lower magnetic fields, reproducing the behavior observed experimentally. When Joule-heating-induced thermal modulation is included, the simulation generates the harmonic Hall signals in response to the fluctuations of the top spin as shown in Fig.~\ref{FIG:4}(b). Large odd harmonics emerge only within the spin-flip region and reach amplitudes comparable to the first harmonic, reproducing the key features of the measurements in Fig.~\ref{FIG:2}. By contrast, simulations including only current-induced spin-orbit torque produce negligible higher-order harmonics (see the SM \cite{SM}), indicating that Joule-heating-induced parametric modulation is the dominant driving mechanism

The microscopic origin of the giant nonlinear response is revealed by the field evolution of the interfacial magnetization shown in Fig.~\ref{FIG:4}(c). The first panel represents the current modulation (blue) and corresponding thermal modulation (red) over two cycles. For fields below $H^*$ (e.g., 1~T), the magnetization remains in the exchange-dominated state throughout the current cycle, hence no harmonic signal generation is expected. In the Zeeman--exchange-frustrated regime, however, the \(2\omega\) temperature modulation periodically changes the balance between Zeeman and exchange energies, driving the magnetization across the stability boundary between the exchange- and Zeeman-dominated states. The resulting switching produces a strongly anharmonic \(m_z(t)\) waveform and, consequently, large higher-order odd harmonics as exemplified in the remaining panels. At fields above $H^{**}$, the Zeeman-dominated state remains stable during the entire cycle and the switching disappears. The simulations therefore identify periodic switching between competing magnetic states, rather than small oscillatory canting about a single equilibrium configuration, as the origin of the giant nonlinear Hall response. 

This mechanism is summarized schematically in the bottom panel of Fig.~\ref{FIG:3}(b). Joule heating periodically modulates the temperature and repeatedly drives the system across a stability boundary in the frustrated regime. The resulting threshold-crossing behavior converts a smooth thermal modulation into a large nonlinear electrical response. Because the modulation originates from Joule heating proportional to \(I^2(t)\), the magnetic response is generated predominantly at even harmonics of the excitation frequency. Mixing these modulations with the \(1\omega\) component of the injected current then produces the observed odd-harmonic Hall voltages, while even harmonics remain negligible (see the SM \cite{SM}).  

Additional measurements shown in the SM \cite{SM} further support this interpretation. Giant nonlinear Hall responses are reproducibly observed in Al:TbIG/Pt heterostructures with different garnet thicknesses and interface conditions, demonstrating that the phenomenon is robust across a broad range of Al:TbIG samples. We also observed additional key features such as clear imaginary components of the Hall signals and a frequency-dependent transition field, both of which further characterize the pronounced nonlinear behavior.

In summary, we have observed a giant nonlinear Hall response in Al:TbIG/Pt arising from Zeeman–exchange frustration near magnetic compensation. A compensation gradient across the film thickness creates competing exchange- and Zeeman-favored magnetic configurations, leading to a field-induced spin-flip instability. We show that Joule-heating-induced thermal modulation periodically drives the interfacial magnetization across this instability, generating odd harmonic Hall voltages with amplitudes comparable to the first harmonic. More broadly, our results identify frustration-assisted switching as a mechanism by which weak periodic perturbations can be converted into giant nonlinear transport responses in compensated magnetic systems.

\textit{Acknowledgments---}
We acknowledge funding from the European Research Council (ERC) under the European Union's Horizon 2020 research and innovation programme (project MAGNEPIC, Grant Agreement No. 949052). T.S. acknowledges the support of the Beatriu de Pin\'os postdoctoral fellowship programme (Grant No. 2023 BP 00168) of the Department of Research and Universities of the Generalitat de Catalunya. 
We thank the scientific services of the Institut de Ci\`encia de Materials de Barcelona (ICMAB-CSIC). This work was conducted in part within the MAI-SKY (PID2021-125973OA-I00) project funded by MCIN/AEI/10.13039/501100011033/FEDER, UE. We acknowledge financial support from the State Investigation Agency, through the Severo Ochoa Programme for Centres of Excellence in R\&D (CEX2023-001263-S).
We acknowledge the use of instrumentation as well as the technical advice provided by the Joint Electron Microscopy Center at ALBA (JEMCA) and funding from Grant IU16-014206 (METCAM-FIB) to ICN2 funded by the European Union through the European Regional Development Fund (ERDF), with the support of the Ministry of Research and Universities, Generalitat de Catalunya. We thank Saúl Estandía for his assistance in replotting the HAADF-STEM and EELS data. 

\bibliography{REFERENCE}

\begin{thebibliography}{32}%
\makeatletter
\providecommand \@ifxundefined [1]{%
 \@ifx{#1\undefined}
}%
\providecommand \@ifnum [1]{%
 \ifnum #1\expandafter \@firstoftwo
 \else \expandafter \@secondoftwo
 \fi
}%
\providecommand \@ifx [1]{%
 \ifx #1\expandafter \@firstoftwo
 \else \expandafter \@secondoftwo
 \fi
}%
\providecommand \natexlab [1]{#1}%
\providecommand \enquote  [1]{``#1''}%
\providecommand \bibnamefont  [1]{#1}%
\providecommand \bibfnamefont [1]{#1}%
\providecommand \citenamefont [1]{#1}%
\providecommand \href@noop [0]{\@secondoftwo}%
\providecommand \href [0]{\begingroup \@sanitize@url \@href}%
\providecommand \@href[1]{\@@startlink{#1}\@@href}%
\providecommand \@@href[1]{\endgroup#1\@@endlink}%
\providecommand \@sanitize@url [0]{\catcode `\\12\catcode `\$12\catcode
  `\&12\catcode `\#12\catcode `\^12\catcode `\_12\catcode `\%12\relax}%
\providecommand \@@startlink[1]{}%
\providecommand \@@endlink[0]{}%
\providecommand \url  [0]{\begingroup\@sanitize@url \@url }%
\providecommand \@url [1]{\endgroup\@href {#1}{\urlprefix }}%
\providecommand \urlprefix  [0]{URL }%
\providecommand \Eprint [0]{\href }%
\providecommand \doibase [0]{https://doi.org/}%
\providecommand \selectlanguage [0]{\@gobble}%
\providecommand \bibinfo  [0]{\@secondoftwo}%
\providecommand \bibfield  [0]{\@secondoftwo}%
\providecommand \translation [1]{[#1]}%
\providecommand \BibitemOpen [0]{}%
\providecommand \bibitemStop [0]{}%
\providecommand \bibitemNoStop [0]{.\EOS\space}%
\providecommand \EOS [0]{\spacefactor3000\relax}%
\providecommand \BibitemShut  [1]{\csname bibitem#1\endcsname}%
\let\auto@bib@innerbib\@empty
\bibitem [{\citenamefont {Zheng}\ \emph {et~al.}(2023)\citenamefont {Zheng},
  \citenamefont {Wang}, \citenamefont {Wang}, \citenamefont {Sun},
  \citenamefont {He}, \citenamefont {Yan},\ and\ \citenamefont
  {Yuan}}]{Zheng2023tutorial}%
  \BibitemOpen
  \bibfield  {author} {\bibinfo {author} {\bibfnamefont {S.}~\bibnamefont
  {Zheng}}, \bibinfo {author} {\bibfnamefont {Z.}~\bibnamefont {Wang}},
  \bibinfo {author} {\bibfnamefont {Y.}~\bibnamefont {Wang}}, \bibinfo {author}
  {\bibfnamefont {F.}~\bibnamefont {Sun}}, \bibinfo {author} {\bibfnamefont
  {Q.}~\bibnamefont {He}}, \bibinfo {author} {\bibfnamefont {P.}~\bibnamefont
  {Yan}},\ and\ \bibinfo {author} {\bibfnamefont {H.}~\bibnamefont {Yuan}},\
  }\bibfield  {title} {\bibinfo {title} {Tutorial: nonlinear magnonics},\
  }\href@noop {} {\bibfield  {journal} {\bibinfo  {journal} {Journal of Applied
  Physics}\ }\textbf {\bibinfo {volume} {134}} (\bibinfo {year}
  {2023})}\BibitemShut {NoStop}%
\bibitem [{\citenamefont {Demokritov}\ \emph {et~al.}(2006)\citenamefont
  {Demokritov}, \citenamefont {Demidov}, \citenamefont {Dzyapko}, \citenamefont
  {Melkov}, \citenamefont {Serga}, \citenamefont {Hillebrands},\ and\
  \citenamefont {Slavin}}]{Demokritov2006bose}%
  \BibitemOpen
  \bibfield  {author} {\bibinfo {author} {\bibfnamefont {S.~O.}\ \bibnamefont
  {Demokritov}}, \bibinfo {author} {\bibfnamefont {V.~E.}\ \bibnamefont
  {Demidov}}, \bibinfo {author} {\bibfnamefont {O.}~\bibnamefont {Dzyapko}},
  \bibinfo {author} {\bibfnamefont {G.~A.}\ \bibnamefont {Melkov}}, \bibinfo
  {author} {\bibfnamefont {A.~A.}\ \bibnamefont {Serga}}, \bibinfo {author}
  {\bibfnamefont {B.}~\bibnamefont {Hillebrands}},\ and\ \bibinfo {author}
  {\bibfnamefont {A.~N.}\ \bibnamefont {Slavin}},\ }\bibfield  {title}
  {\bibinfo {title} {Bose--einstein condensation of quasi-equilibrium magnons
  at room temperature under pumping},\ }\href@noop {} {\bibfield  {journal}
  {\bibinfo  {journal} {Nature}\ }\textbf {\bibinfo {volume} {443}},\ \bibinfo
  {pages} {430} (\bibinfo {year} {2006})}\BibitemShut {NoStop}%
\bibitem [{\citenamefont {Kiselev}\ \emph {et~al.}(2003)\citenamefont
  {Kiselev}, \citenamefont {Sankey}, \citenamefont {Krivorotov}, \citenamefont
  {Emley}, \citenamefont {Schoelkopf}, \citenamefont {Buhrman},\ and\
  \citenamefont {Ralph}}]{Kiselev2003microwave}%
  \BibitemOpen
  \bibfield  {author} {\bibinfo {author} {\bibfnamefont {S.~I.}\ \bibnamefont
  {Kiselev}}, \bibinfo {author} {\bibfnamefont {J.}~\bibnamefont {Sankey}},
  \bibinfo {author} {\bibfnamefont {I.}~\bibnamefont {Krivorotov}}, \bibinfo
  {author} {\bibfnamefont {N.}~\bibnamefont {Emley}}, \bibinfo {author}
  {\bibfnamefont {R.}~\bibnamefont {Schoelkopf}}, \bibinfo {author}
  {\bibfnamefont {R.}~\bibnamefont {Buhrman}},\ and\ \bibinfo {author}
  {\bibfnamefont {D.}~\bibnamefont {Ralph}},\ }\bibfield  {title} {\bibinfo
  {title} {Microwave oscillations of a nanomagnet driven by a spin-polarized
  current},\ }\href@noop {} {\bibfield  {journal} {\bibinfo  {journal}
  {nature}\ }\textbf {\bibinfo {volume} {425}},\ \bibinfo {pages} {380}
  (\bibinfo {year} {2003})}\BibitemShut {NoStop}%
\bibitem [{\citenamefont {Demidov}\ \emph {et~al.}(2012)\citenamefont
  {Demidov}, \citenamefont {Urazhdin}, \citenamefont {Ulrichs}, \citenamefont
  {Tiberkevich}, \citenamefont {Slavin}, \citenamefont {Baither}, \citenamefont
  {Schmitz},\ and\ \citenamefont {Demokritov}}]{Demidov2012magnetic}%
  \BibitemOpen
  \bibfield  {author} {\bibinfo {author} {\bibfnamefont {V.~E.}\ \bibnamefont
  {Demidov}}, \bibinfo {author} {\bibfnamefont {S.}~\bibnamefont {Urazhdin}},
  \bibinfo {author} {\bibfnamefont {H.}~\bibnamefont {Ulrichs}}, \bibinfo
  {author} {\bibfnamefont {V.}~\bibnamefont {Tiberkevich}}, \bibinfo {author}
  {\bibfnamefont {A.}~\bibnamefont {Slavin}}, \bibinfo {author} {\bibfnamefont
  {D.}~\bibnamefont {Baither}}, \bibinfo {author} {\bibfnamefont
  {G.}~\bibnamefont {Schmitz}},\ and\ \bibinfo {author} {\bibfnamefont {S.~O.}\
  \bibnamefont {Demokritov}},\ }\bibfield  {title} {\bibinfo {title} {Magnetic
  nano-oscillator driven by pure spin current},\ }\href@noop {} {\bibfield
  {journal} {\bibinfo  {journal} {Nature materials}\ }\textbf {\bibinfo
  {volume} {11}},\ \bibinfo {pages} {1028} (\bibinfo {year}
  {2012})}\BibitemShut {NoStop}%
\bibitem [{\citenamefont {Lerose}\ \emph {et~al.}(2019)\citenamefont {Lerose},
  \citenamefont {Marino}, \citenamefont {Gambassi},\ and\ \citenamefont
  {Silva}}]{Lerose2019prethermal}%
  \BibitemOpen
  \bibfield  {author} {\bibinfo {author} {\bibfnamefont {A.}~\bibnamefont
  {Lerose}}, \bibinfo {author} {\bibfnamefont {J.}~\bibnamefont {Marino}},
  \bibinfo {author} {\bibfnamefont {A.}~\bibnamefont {Gambassi}},\ and\
  \bibinfo {author} {\bibfnamefont {A.}~\bibnamefont {Silva}},\ }\bibfield
  {title} {\bibinfo {title} {Prethermal quantum many-body kapitza phases of
  periodically driven spin systems},\ }\href@noop {} {\bibfield  {journal}
  {\bibinfo  {journal} {Physical Review B}\ }\textbf {\bibinfo {volume}
  {100}},\ \bibinfo {pages} {104306} (\bibinfo {year} {2019})}\BibitemShut
  {NoStop}%
\bibitem [{\citenamefont {Kulikov}\ \emph {et~al.}(2022)\citenamefont
  {Kulikov}, \citenamefont {Anghel}, \citenamefont {Preda}, \citenamefont
  {Nashaat}, \citenamefont {Sameh},\ and\ \citenamefont
  {Shukrinov}}]{Kulikov2022kapitza}%
  \BibitemOpen
  \bibfield  {author} {\bibinfo {author} {\bibfnamefont {K.}~\bibnamefont
  {Kulikov}}, \bibinfo {author} {\bibfnamefont {D.}~\bibnamefont {Anghel}},
  \bibinfo {author} {\bibfnamefont {A.}~\bibnamefont {Preda}}, \bibinfo
  {author} {\bibfnamefont {M.}~\bibnamefont {Nashaat}}, \bibinfo {author}
  {\bibfnamefont {M.}~\bibnamefont {Sameh}},\ and\ \bibinfo {author}
  {\bibfnamefont {Y.~M.}\ \bibnamefont {Shukrinov}},\ }\bibfield  {title}
  {\bibinfo {title} {Kapitza pendulum effects in a josephson junction coupled
  to a nanomagnet under external periodic drive},\ }\href@noop {} {\bibfield
  {journal} {\bibinfo  {journal} {Physical Review B}\ }\textbf {\bibinfo
  {volume} {105}},\ \bibinfo {pages} {094421} (\bibinfo {year}
  {2022})}\BibitemShut {NoStop}%
\bibitem [{\citenamefont {Kurebayashi}\ \emph {et~al.}(2026)\citenamefont
  {Kurebayashi}, \citenamefont {Barker}, \citenamefont {Yamazaki},
  \citenamefont {Kushwaha}, \citenamefont {Stenning}, \citenamefont {Youel},
  \citenamefont {Hou}, \citenamefont {Dion}, \citenamefont {Prestwood},
  \citenamefont {Bauer} \emph {et~al.}}]{Kurebayashi2026dynamical}%
  \BibitemOpen
  \bibfield  {author} {\bibinfo {author} {\bibfnamefont {H.}~\bibnamefont
  {Kurebayashi}}, \bibinfo {author} {\bibfnamefont {J.}~\bibnamefont {Barker}},
  \bibinfo {author} {\bibfnamefont {T.}~\bibnamefont {Yamazaki}}, \bibinfo
  {author} {\bibfnamefont {V.~K.}\ \bibnamefont {Kushwaha}}, \bibinfo {author}
  {\bibfnamefont {K.~D.}\ \bibnamefont {Stenning}}, \bibinfo {author}
  {\bibfnamefont {H.}~\bibnamefont {Youel}}, \bibinfo {author} {\bibfnamefont
  {X.}~\bibnamefont {Hou}}, \bibinfo {author} {\bibfnamefont {T.}~\bibnamefont
  {Dion}}, \bibinfo {author} {\bibfnamefont {D.}~\bibnamefont {Prestwood}},
  \bibinfo {author} {\bibfnamefont {G.~E.}\ \bibnamefont {Bauer}}, \emph
  {et~al.},\ }\bibfield  {title} {\bibinfo {title} {Dynamical stability by spin
  transfer in nearly isotropic magnets},\ }\href@noop {} {\bibfield  {journal}
  {\bibinfo  {journal} {Nature Materials}\ ,\ \bibinfo {pages} {1}} (\bibinfo
  {year} {2026})}\BibitemShut {NoStop}%
\bibitem [{\citenamefont {Garello}\ \emph {et~al.}(2013)\citenamefont
  {Garello}, \citenamefont {Miron}, \citenamefont {Avci}, \citenamefont
  {Freimuth}, \citenamefont {Mokrousov}, \citenamefont {Bl{\"u}gel},
  \citenamefont {Auffret}, \citenamefont {Boulle}, \citenamefont {Gaudin},\
  and\ \citenamefont {Gambardella}}]{garello2013symmetry}%
  \BibitemOpen
  \bibfield  {author} {\bibinfo {author} {\bibfnamefont {K.}~\bibnamefont
  {Garello}}, \bibinfo {author} {\bibfnamefont {I.~M.}\ \bibnamefont {Miron}},
  \bibinfo {author} {\bibfnamefont {C.~O.}\ \bibnamefont {Avci}}, \bibinfo
  {author} {\bibfnamefont {F.}~\bibnamefont {Freimuth}}, \bibinfo {author}
  {\bibfnamefont {Y.}~\bibnamefont {Mokrousov}}, \bibinfo {author}
  {\bibfnamefont {S.}~\bibnamefont {Bl{\"u}gel}}, \bibinfo {author}
  {\bibfnamefont {S.}~\bibnamefont {Auffret}}, \bibinfo {author} {\bibfnamefont
  {O.}~\bibnamefont {Boulle}}, \bibinfo {author} {\bibfnamefont
  {G.}~\bibnamefont {Gaudin}},\ and\ \bibinfo {author} {\bibfnamefont
  {P.}~\bibnamefont {Gambardella}},\ }\bibfield  {title} {\bibinfo {title}
  {Symmetry and magnitude of spin--orbit torques in ferromagnetic
  heterostructures},\ }\href@noop {} {\bibfield  {journal} {\bibinfo  {journal}
  {Nature nanotechnology}\ }\textbf {\bibinfo {volume} {8}},\ \bibinfo {pages}
  {587} (\bibinfo {year} {2013})}\BibitemShut {NoStop}%
\bibitem [{\citenamefont {Avci}\ \emph {et~al.}(2015)\citenamefont {Avci},
  \citenamefont {Garello}, \citenamefont {Ghosh}, \citenamefont {Gabureac},
  \citenamefont {Alvarado},\ and\ \citenamefont
  {Gambardella}}]{Avci2015unidirectional}%
  \BibitemOpen
  \bibfield  {author} {\bibinfo {author} {\bibfnamefont {C.~O.}\ \bibnamefont
  {Avci}}, \bibinfo {author} {\bibfnamefont {K.}~\bibnamefont {Garello}},
  \bibinfo {author} {\bibfnamefont {A.}~\bibnamefont {Ghosh}}, \bibinfo
  {author} {\bibfnamefont {M.}~\bibnamefont {Gabureac}}, \bibinfo {author}
  {\bibfnamefont {S.~F.}\ \bibnamefont {Alvarado}},\ and\ \bibinfo {author}
  {\bibfnamefont {P.}~\bibnamefont {Gambardella}},\ }\bibfield  {title}
  {\bibinfo {title} {Unidirectional spin hall magnetoresistance in
  ferromagnet/normal metal bilayers},\ }\href@noop {} {\bibfield  {journal}
  {\bibinfo  {journal} {Nature Physics}\ }\textbf {\bibinfo {volume} {11}},\
  \bibinfo {pages} {570} (\bibinfo {year} {2015})}\BibitemShut {NoStop}%
\bibitem [{\citenamefont {He}\ \emph {et~al.}(2018)\citenamefont {He},
  \citenamefont {Zhang}, \citenamefont {Zhu}, \citenamefont {Liu},
  \citenamefont {Wang}, \citenamefont {Yu}, \citenamefont {Vignale},\ and\
  \citenamefont {Yang}}]{he2018bilinear}%
  \BibitemOpen
  \bibfield  {author} {\bibinfo {author} {\bibfnamefont {P.}~\bibnamefont
  {He}}, \bibinfo {author} {\bibfnamefont {S.~S.-L.}\ \bibnamefont {Zhang}},
  \bibinfo {author} {\bibfnamefont {D.}~\bibnamefont {Zhu}}, \bibinfo {author}
  {\bibfnamefont {Y.}~\bibnamefont {Liu}}, \bibinfo {author} {\bibfnamefont
  {Y.}~\bibnamefont {Wang}}, \bibinfo {author} {\bibfnamefont {J.}~\bibnamefont
  {Yu}}, \bibinfo {author} {\bibfnamefont {G.}~\bibnamefont {Vignale}},\ and\
  \bibinfo {author} {\bibfnamefont {H.}~\bibnamefont {Yang}},\ }\bibfield
  {title} {\bibinfo {title} {Bilinear magnetoelectric resistance as a probe of
  three-dimensional spin texture in topological surface states},\ }\href@noop
  {} {\bibfield  {journal} {\bibinfo  {journal} {Nature Physics}\ }\textbf
  {\bibinfo {volume} {14}},\ \bibinfo {pages} {495} (\bibinfo {year}
  {2018})}\BibitemShut {NoStop}%
\bibitem [{\citenamefont {Cheng}\ \emph {et~al.}(2022)\citenamefont {Cheng},
  \citenamefont {Cogulu}, \citenamefont {Resnick}, \citenamefont {Michel},
  \citenamefont {Statuto}, \citenamefont {Kent},\ and\ \citenamefont
  {Yang}}]{Cheng2022third}%
  \BibitemOpen
  \bibfield  {author} {\bibinfo {author} {\bibfnamefont {Y.}~\bibnamefont
  {Cheng}}, \bibinfo {author} {\bibfnamefont {E.}~\bibnamefont {Cogulu}},
  \bibinfo {author} {\bibfnamefont {R.~D.}\ \bibnamefont {Resnick}}, \bibinfo
  {author} {\bibfnamefont {J.~J.}\ \bibnamefont {Michel}}, \bibinfo {author}
  {\bibfnamefont {N.~N.}\ \bibnamefont {Statuto}}, \bibinfo {author}
  {\bibfnamefont {A.~D.}\ \bibnamefont {Kent}},\ and\ \bibinfo {author}
  {\bibfnamefont {F.}~\bibnamefont {Yang}},\ }\bibfield  {title} {\bibinfo
  {title} {Third harmonic characterization of antiferromagnetic
  heterostructures},\ }\href@noop {} {\bibfield  {journal} {\bibinfo  {journal}
  {Nature communications}\ }\textbf {\bibinfo {volume} {13}},\ \bibinfo {pages}
  {3659} (\bibinfo {year} {2022})}\BibitemShut {NoStop}%
\bibitem [{\citenamefont {N{\'e}el}(1972)}]{Neel1972}%
  \BibitemOpen
  \bibfield  {author} {\bibinfo {author} {\bibfnamefont {L.}~\bibnamefont
  {N{\'e}el}},\ }\bibfield  {title} {\bibinfo {title} {Magnetism and the local
  molecular field},\ }\href@noop {} {\bibfield  {journal} {\bibinfo  {journal}
  {Nobel Lectures Physics 1963-1970}\ } (\bibinfo {year} {1972})}\BibitemShut
  {NoStop}%
\bibitem [{\citenamefont {Dionne}(2009)}]{Dionne2009}%
  \BibitemOpen
  \bibfield  {author} {\bibinfo {author} {\bibfnamefont {G.~F.}\ \bibnamefont
  {Dionne}},\ }\bibfield  {title} {\bibinfo {title} {Magnetic oxides},\ }\href
  {https://doi.org/10.1007/978-1-4419-0054-8/COVER} {\bibfield  {journal}
  {\bibinfo  {journal} {Magnetic Oxides}\ ,\ \bibinfo {pages} {1}} (\bibinfo
  {year} {2009})}\BibitemShut {NoStop}%
\bibitem [{\citenamefont {Lahoubi}\ \emph {et~al.}(2003)\citenamefont
  {Lahoubi}, \citenamefont {Guillot}, \citenamefont {Marchand}, \citenamefont
  {Tcheou},\ and\ \citenamefont {Roudault}}]{Lahoubi2003double}%
  \BibitemOpen
  \bibfield  {author} {\bibinfo {author} {\bibfnamefont {M.}~\bibnamefont
  {Lahoubi}}, \bibinfo {author} {\bibfnamefont {M.}~\bibnamefont {Guillot}},
  \bibinfo {author} {\bibfnamefont {A.}~\bibnamefont {Marchand}}, \bibinfo
  {author} {\bibfnamefont {F.}~\bibnamefont {Tcheou}},\ and\ \bibinfo {author}
  {\bibfnamefont {E.}~\bibnamefont {Roudault}},\ }\bibfield  {title} {\bibinfo
  {title} {Double umbrella structure in terbium iron garnet},\ }\href@noop {}
  {\bibfield  {journal} {\bibinfo  {journal} {IEEE Transactions on Magnetics}\
  }\textbf {\bibinfo {volume} {20}},\ \bibinfo {pages} {1518} (\bibinfo {year}
  {2003})}\BibitemShut {NoStop}%
\bibitem [{\citenamefont {Li}\ \emph {et~al.}(2024)\citenamefont {Li},
  \citenamefont {Duan}, \citenamefont {Wang}, \citenamefont {Lang},
  \citenamefont {Zhang}, \citenamefont {Yang}, \citenamefont {Li},
  \citenamefont {Fan}, \citenamefont {Shen}, \citenamefont {Shi} \emph
  {et~al.}}]{Li2024giant}%
  \BibitemOpen
  \bibfield  {author} {\bibinfo {author} {\bibfnamefont {Y.}~\bibnamefont
  {Li}}, \bibinfo {author} {\bibfnamefont {Y.}~\bibnamefont {Duan}}, \bibinfo
  {author} {\bibfnamefont {M.}~\bibnamefont {Wang}}, \bibinfo {author}
  {\bibfnamefont {L.}~\bibnamefont {Lang}}, \bibinfo {author} {\bibfnamefont
  {Y.}~\bibnamefont {Zhang}}, \bibinfo {author} {\bibfnamefont
  {M.}~\bibnamefont {Yang}}, \bibinfo {author} {\bibfnamefont {J.}~\bibnamefont
  {Li}}, \bibinfo {author} {\bibfnamefont {W.}~\bibnamefont {Fan}}, \bibinfo
  {author} {\bibfnamefont {K.}~\bibnamefont {Shen}}, \bibinfo {author}
  {\bibfnamefont {Z.}~\bibnamefont {Shi}}, \emph {et~al.},\ }\bibfield  {title}
  {\bibinfo {title} {Giant magnon-polaron anomalies in spin seebeck effect in
  double umbrella-structured tb 3 fe 5 o 12 films},\ }\href@noop {} {\bibfield
  {journal} {\bibinfo  {journal} {Physical Review Letters}\ }\textbf {\bibinfo
  {volume} {132}},\ \bibinfo {pages} {056702} (\bibinfo {year}
  {2024})}\BibitemShut {NoStop}%
\bibitem [{\citenamefont {Kirilyuk}\ \emph {et~al.}(2010)\citenamefont
  {Kirilyuk}, \citenamefont {Kimel},\ and\ \citenamefont
  {Rasing}}]{Kirilyuk2010ultrafast}%
  \BibitemOpen
  \bibfield  {author} {\bibinfo {author} {\bibfnamefont {A.}~\bibnamefont
  {Kirilyuk}}, \bibinfo {author} {\bibfnamefont {A.~V.}\ \bibnamefont
  {Kimel}},\ and\ \bibinfo {author} {\bibfnamefont {T.}~\bibnamefont
  {Rasing}},\ }\bibfield  {title} {\bibinfo {title} {Ultrafast optical
  manipulation of magnetic order},\ }\href@noop {} {\bibfield  {journal}
  {\bibinfo  {journal} {Reviews of Modern Physics}\ }\textbf {\bibinfo {volume}
  {82}},\ \bibinfo {pages} {2731} (\bibinfo {year} {2010})}\BibitemShut
  {NoStop}%
\bibitem [{\citenamefont {Kim}\ \emph {et~al.}(2017)\citenamefont {Kim},
  \citenamefont {Kim}, \citenamefont {Hirata}, \citenamefont {Oh},
  \citenamefont {Tono}, \citenamefont {Kim}, \citenamefont {Okuno},
  \citenamefont {Ham}, \citenamefont {Kim}, \citenamefont {Go}, \citenamefont
  {Tserkovnyak}, \citenamefont {Tsukamoto}, \citenamefont {Moriyama},
  \citenamefont {Lee},\ and\ \citenamefont {Ono}}]{Kim2017}%
  \BibitemOpen
  \bibfield  {author} {\bibinfo {author} {\bibfnamefont {K.-J.}\ \bibnamefont
  {Kim}}, \bibinfo {author} {\bibfnamefont {S.~K.}\ \bibnamefont {Kim}},
  \bibinfo {author} {\bibfnamefont {Y.}~\bibnamefont {Hirata}}, \bibinfo
  {author} {\bibfnamefont {S.-H.}\ \bibnamefont {Oh}}, \bibinfo {author}
  {\bibfnamefont {T.}~\bibnamefont {Tono}}, \bibinfo {author} {\bibfnamefont
  {D.-H.}\ \bibnamefont {Kim}}, \bibinfo {author} {\bibfnamefont
  {T.}~\bibnamefont {Okuno}}, \bibinfo {author} {\bibfnamefont {W.~S.}\
  \bibnamefont {Ham}}, \bibinfo {author} {\bibfnamefont {S.}~\bibnamefont
  {Kim}}, \bibinfo {author} {\bibfnamefont {G.}~\bibnamefont {Go}}, \bibinfo
  {author} {\bibfnamefont {Y.}~\bibnamefont {Tserkovnyak}}, \bibinfo {author}
  {\bibfnamefont {A.}~\bibnamefont {Tsukamoto}}, \bibinfo {author}
  {\bibfnamefont {T.}~\bibnamefont {Moriyama}}, \bibinfo {author}
  {\bibfnamefont {K.-J.}\ \bibnamefont {Lee}},\ and\ \bibinfo {author}
  {\bibfnamefont {T.}~\bibnamefont {Ono}},\ }\bibfield  {title} {\bibinfo
  {title} {Fast domain wall motion in the vicinity of the angular momentum
  compensation temperature of ferrimagnets},\ }\href
  {https://doi.org/10.1038/nmat4990} {\bibfield  {journal} {\bibinfo  {journal}
  {Nature Materials}\ }\textbf {\bibinfo {volume} {16}},\ \bibinfo {pages}
  {1187} (\bibinfo {year} {2017})}\BibitemShut {NoStop}%
\bibitem [{\citenamefont {Caretta}\ \emph {et~al.}(2018)\citenamefont
  {Caretta}, \citenamefont {Mann}, \citenamefont {Büttner}, \citenamefont
  {Ueda}, \citenamefont {Pfau}, \citenamefont {Günther}, \citenamefont
  {Hessing}, \citenamefont {Churikova}, \citenamefont {Klose}, \citenamefont
  {Schneider}, \citenamefont {Engel}, \citenamefont {Marcus}, \citenamefont
  {Bono}, \citenamefont {Bagschik}, \citenamefont {Eisebitt},\ and\
  \citenamefont {Beach}}]{Caretta2018}%
  \BibitemOpen
  \bibfield  {author} {\bibinfo {author} {\bibfnamefont {L.}~\bibnamefont
  {Caretta}}, \bibinfo {author} {\bibfnamefont {M.}~\bibnamefont {Mann}},
  \bibinfo {author} {\bibfnamefont {F.}~\bibnamefont {Büttner}}, \bibinfo
  {author} {\bibfnamefont {K.}~\bibnamefont {Ueda}}, \bibinfo {author}
  {\bibfnamefont {B.}~\bibnamefont {Pfau}}, \bibinfo {author} {\bibfnamefont
  {C.~M.}\ \bibnamefont {Günther}}, \bibinfo {author} {\bibfnamefont
  {P.}~\bibnamefont {Hessing}}, \bibinfo {author} {\bibfnamefont
  {A.}~\bibnamefont {Churikova}}, \bibinfo {author} {\bibfnamefont
  {C.}~\bibnamefont {Klose}}, \bibinfo {author} {\bibfnamefont
  {M.}~\bibnamefont {Schneider}}, \bibinfo {author} {\bibfnamefont
  {D.}~\bibnamefont {Engel}}, \bibinfo {author} {\bibfnamefont
  {C.}~\bibnamefont {Marcus}}, \bibinfo {author} {\bibfnamefont
  {D.}~\bibnamefont {Bono}}, \bibinfo {author} {\bibfnamefont {K.}~\bibnamefont
  {Bagschik}}, \bibinfo {author} {\bibfnamefont {S.}~\bibnamefont {Eisebitt}},\
  and\ \bibinfo {author} {\bibfnamefont {G.~S.~D.}\ \bibnamefont {Beach}},\
  }\bibfield  {title} {\bibinfo {title} {Fast current-driven domain walls and
  small skyrmions in a compensated ferrimagnet},\ }\href
  {https://doi.org/10.1038/s41565-018-0255-3} {\bibfield  {journal} {\bibinfo
  {journal} {Nature Nanotechnology}\ }\textbf {\bibinfo {volume} {13}},\
  \bibinfo {pages} {1154} (\bibinfo {year} {2018})}\BibitemShut {NoStop}%
\bibitem [{\citenamefont {Liensberger}\ \emph {et~al.}(2019)\citenamefont
  {Liensberger}, \citenamefont {Kamra}, \citenamefont {Maier-Flaig},
  \citenamefont {Gepr{\"a}gs}, \citenamefont {Erb}, \citenamefont
  {Goennenwein}, \citenamefont {Gross}, \citenamefont {Belzig}, \citenamefont
  {Huebl},\ and\ \citenamefont {Weiler}}]{Liensberger2019exchange}%
  \BibitemOpen
  \bibfield  {author} {\bibinfo {author} {\bibfnamefont {L.}~\bibnamefont
  {Liensberger}}, \bibinfo {author} {\bibfnamefont {A.}~\bibnamefont {Kamra}},
  \bibinfo {author} {\bibfnamefont {H.}~\bibnamefont {Maier-Flaig}}, \bibinfo
  {author} {\bibfnamefont {S.}~\bibnamefont {Gepr{\"a}gs}}, \bibinfo {author}
  {\bibfnamefont {A.}~\bibnamefont {Erb}}, \bibinfo {author} {\bibfnamefont
  {S.~T.}\ \bibnamefont {Goennenwein}}, \bibinfo {author} {\bibfnamefont
  {R.}~\bibnamefont {Gross}}, \bibinfo {author} {\bibfnamefont
  {W.}~\bibnamefont {Belzig}}, \bibinfo {author} {\bibfnamefont
  {H.}~\bibnamefont {Huebl}},\ and\ \bibinfo {author} {\bibfnamefont
  {M.}~\bibnamefont {Weiler}},\ }\bibfield  {title} {\bibinfo {title}
  {Exchange-enhanced ultrastrong magnon-magnon coupling in a compensated
  ferrimagnet},\ }\href@noop {} {\bibfield  {journal} {\bibinfo  {journal}
  {Physical review letters}\ }\textbf {\bibinfo {volume} {123}},\ \bibinfo
  {pages} {117204} (\bibinfo {year} {2019})}\BibitemShut {NoStop}%
\bibitem [{\citenamefont {Deb}\ \emph {et~al.}(2021)\citenamefont {Deb},
  \citenamefont {Molho},\ and\ \citenamefont {Barbara}}]{Deb2021tunable}%
  \BibitemOpen
  \bibfield  {author} {\bibinfo {author} {\bibfnamefont {M.}~\bibnamefont
  {Deb}}, \bibinfo {author} {\bibfnamefont {P.}~\bibnamefont {Molho}},\ and\
  \bibinfo {author} {\bibfnamefont {B.}~\bibnamefont {Barbara}},\ }\bibfield
  {title} {\bibinfo {title} {Tunable exchange-bias-like effect in
  bi-substituted gadolinium iron garnet film},\ }\href@noop {} {\bibfield
  {journal} {\bibinfo  {journal} {Physical Review Applied}\ }\textbf {\bibinfo
  {volume} {15}},\ \bibinfo {pages} {054064} (\bibinfo {year}
  {2021})}\BibitemShut {NoStop}%
\bibitem [{\citenamefont {Wang}\ \emph {et~al.}(2021)\citenamefont {Wang},
  \citenamefont {Wang}, \citenamefont {Clark}, \citenamefont {Chen},
  \citenamefont {Cheng}, \citenamefont {Freeland},\ and\ \citenamefont
  {Xiao}}]{Wang2021probing}%
  \BibitemOpen
  \bibfield  {author} {\bibinfo {author} {\bibfnamefont {Y.}~\bibnamefont
  {Wang}}, \bibinfo {author} {\bibfnamefont {X.}~\bibnamefont {Wang}}, \bibinfo
  {author} {\bibfnamefont {A.~T.}\ \bibnamefont {Clark}}, \bibinfo {author}
  {\bibfnamefont {H.}~\bibnamefont {Chen}}, \bibinfo {author} {\bibfnamefont
  {X.~M.}\ \bibnamefont {Cheng}}, \bibinfo {author} {\bibfnamefont {J.~W.}\
  \bibnamefont {Freeland}},\ and\ \bibinfo {author} {\bibfnamefont {J.~Q.}\
  \bibnamefont {Xiao}},\ }\bibfield  {title} {\bibinfo {title} {Probing
  exchange bias at the surface of a doped ferrimagnetic insulator},\
  }\href@noop {} {\bibfield  {journal} {\bibinfo  {journal} {Physical Review
  Materials}\ }\textbf {\bibinfo {volume} {5}},\ \bibinfo {pages} {074409}
  (\bibinfo {year} {2021})}\BibitemShut {NoStop}%
\bibitem [{\citenamefont {Shiino}\ \emph {et~al.}(2025)\citenamefont {Shiino},
  \citenamefont {Fettizio}, \citenamefont {Estand{\'\i}a},\ and\ \citenamefont
  {Avci}}]{Shiino2025tunable}%
  \BibitemOpen
  \bibfield  {author} {\bibinfo {author} {\bibfnamefont {T.}~\bibnamefont
  {Shiino}}, \bibinfo {author} {\bibfnamefont {M.}~\bibnamefont {Fettizio}},
  \bibinfo {author} {\bibfnamefont {S.}~\bibnamefont {Estand{\'\i}a}},\ and\
  \bibinfo {author} {\bibfnamefont {C.~O.}\ \bibnamefont {Avci}},\ }\bibfield
  {title} {\bibinfo {title} {Tunable magnetism and intrinsic exchange bias in
  al-substituted terbium iron garnet},\ }\href@noop {} {\bibfield  {journal}
  {\bibinfo  {journal} {Advanced Materials}\ ,\ \bibinfo {pages} {e10669}}
  (\bibinfo {year} {2025})}\BibitemShut {NoStop}%
\bibitem [{\citenamefont {Nakayama}\ \emph {et~al.}(2013)\citenamefont
  {Nakayama}, \citenamefont {Althammer}, \citenamefont {Chen}, \citenamefont
  {Uchida}, \citenamefont {Kajiwara}, \citenamefont {Kikuchi}, \citenamefont
  {Ohtani}, \citenamefont {Gepr\"ags}, \citenamefont {Opel}, \citenamefont
  {Takahashi}, \citenamefont {Gross}, \citenamefont {Bauer}, \citenamefont
  {Goennenwein},\ and\ \citenamefont {Saitoh}}]{Nakayama2013}%
  \BibitemOpen
  \bibfield  {author} {\bibinfo {author} {\bibfnamefont {H.}~\bibnamefont
  {Nakayama}}, \bibinfo {author} {\bibfnamefont {M.}~\bibnamefont {Althammer}},
  \bibinfo {author} {\bibfnamefont {Y.-T.}\ \bibnamefont {Chen}}, \bibinfo
  {author} {\bibfnamefont {K.}~\bibnamefont {Uchida}}, \bibinfo {author}
  {\bibfnamefont {Y.}~\bibnamefont {Kajiwara}}, \bibinfo {author}
  {\bibfnamefont {D.}~\bibnamefont {Kikuchi}}, \bibinfo {author} {\bibfnamefont
  {T.}~\bibnamefont {Ohtani}}, \bibinfo {author} {\bibfnamefont
  {S.}~\bibnamefont {Gepr\"ags}}, \bibinfo {author} {\bibfnamefont
  {M.}~\bibnamefont {Opel}}, \bibinfo {author} {\bibfnamefont {S.}~\bibnamefont
  {Takahashi}}, \bibinfo {author} {\bibfnamefont {R.}~\bibnamefont {Gross}},
  \bibinfo {author} {\bibfnamefont {G.~E.~W.}\ \bibnamefont {Bauer}}, \bibinfo
  {author} {\bibfnamefont {S.~T.~B.}\ \bibnamefont {Goennenwein}},\ and\
  \bibinfo {author} {\bibfnamefont {E.}~\bibnamefont {Saitoh}},\ }\bibfield
  {title} {\bibinfo {title} {Spin hall magnetoresistance induced by a
  nonequilibrium proximity effect},\ }\href
  {https://doi.org/10.1103/PhysRevLett.110.206601} {\bibfield  {journal}
  {\bibinfo  {journal} {Phys. Rev. Lett.}\ }\textbf {\bibinfo {volume} {110}},\
  \bibinfo {pages} {206601} (\bibinfo {year} {2013})}\BibitemShut {NoStop}%
\bibitem [{\citenamefont {Avci}\ \emph {et~al.}(2017)\citenamefont {Avci},
  \citenamefont {Quindeau}, \citenamefont {Pai}, \citenamefont {Mann},
  \citenamefont {Caretta}, \citenamefont {Tang}, \citenamefont {Onbasli},
  \citenamefont {Ross},\ and\ \citenamefont {Beach}}]{avci2017NM}%
  \BibitemOpen
  \bibfield  {author} {\bibinfo {author} {\bibfnamefont {C.~O.}\ \bibnamefont
  {Avci}}, \bibinfo {author} {\bibfnamefont {A.}~\bibnamefont {Quindeau}},
  \bibinfo {author} {\bibfnamefont {C.-F.}\ \bibnamefont {Pai}}, \bibinfo
  {author} {\bibfnamefont {M.}~\bibnamefont {Mann}}, \bibinfo {author}
  {\bibfnamefont {L.}~\bibnamefont {Caretta}}, \bibinfo {author} {\bibfnamefont
  {A.~S.}\ \bibnamefont {Tang}}, \bibinfo {author} {\bibfnamefont {M.~C.}\
  \bibnamefont {Onbasli}}, \bibinfo {author} {\bibfnamefont {C.~A.}\
  \bibnamefont {Ross}},\ and\ \bibinfo {author} {\bibfnamefont {G.~S.}\
  \bibnamefont {Beach}},\ }\bibfield  {title} {\bibinfo {title}
  {Current-induced switching in a magnetic insulator},\ }\href@noop {}
  {\bibfield  {journal} {\bibinfo  {journal} {Nature materials}\ }\textbf
  {\bibinfo {volume} {16}},\ \bibinfo {pages} {309} (\bibinfo {year}
  {2017})}\BibitemShut {NoStop}%
\bibitem [{SM()}]{SM}%
  \BibitemOpen
  \href@noop {} {}\bibinfo {note} {See Supplemental Material at [URL will be
  inserted by publisher] for experimental conditions, additional experimental
  data, simulation conditions, and supplementary simulation results, which
  includes Refs. \cite{fedel2025AFM, song2024temperature, Shiino2025tunable,
  Low2013, malozemoff2013magnetic, liu2011spin}.}\BibitemShut {Stop}%
\bibitem [{\citenamefont {Momma}\ and\ \citenamefont
  {Izumi}(2011)}]{Momma2011vesta}%
  \BibitemOpen
  \bibfield  {author} {\bibinfo {author} {\bibfnamefont {K.}~\bibnamefont
  {Momma}}\ and\ \bibinfo {author} {\bibfnamefont {F.}~\bibnamefont {Izumi}},\
  }\bibfield  {title} {\bibinfo {title} {Vesta 3 for three-dimensional
  visualization of crystal, volumetric and morphology data},\ }\href@noop {}
  {\bibfield  {journal} {\bibinfo  {journal} {Applied Crystallography}\
  }\textbf {\bibinfo {volume} {44}},\ \bibinfo {pages} {1272} (\bibinfo {year}
  {2011})}\BibitemShut {NoStop}%
\bibitem [{\citenamefont {Song}\ \emph {et~al.}(2024)\citenamefont {Song},
  \citenamefont {Lasinger}, \citenamefont {Tang}, \citenamefont {Li},
  \citenamefont {Beach},\ and\ \citenamefont {Ross}}]{song2024temperature}%
  \BibitemOpen
  \bibfield  {author} {\bibinfo {author} {\bibfnamefont {Y.}~\bibnamefont
  {Song}}, \bibinfo {author} {\bibfnamefont {K.}~\bibnamefont {Lasinger}},
  \bibinfo {author} {\bibfnamefont {H.}~\bibnamefont {Tang}}, \bibinfo {author}
  {\bibfnamefont {J.}~\bibnamefont {Li}}, \bibinfo {author} {\bibfnamefont
  {G.~S.}\ \bibnamefont {Beach}},\ and\ \bibinfo {author} {\bibfnamefont
  {C.~A.}\ \bibnamefont {Ross}},\ }\bibfield  {title} {\bibinfo {title}
  {Temperature-dependent surface anisotropy in (110) epitaxial rare earth iron
  garnet films},\ }\href@noop {} {\bibfield  {journal} {\bibinfo  {journal}
  {Small}\ }\textbf {\bibinfo {volume} {20}},\ \bibinfo {pages} {2407381}
  (\bibinfo {year} {2024})}\BibitemShut {NoStop}%
\bibitem [{\citenamefont {Feng}\ \emph {et~al.}(2026)\citenamefont {Feng},
  \citenamefont {Chen}, \citenamefont {Wu}, \citenamefont {Wu}, \citenamefont
  {Li}, \citenamefont {Jiang},\ and\ \citenamefont {Wu}}]{Feng2026}%
  \BibitemOpen
  \bibfield  {author} {\bibinfo {author} {\bibfnamefont {Y.}~\bibnamefont
  {Feng}}, \bibinfo {author} {\bibfnamefont {H.}~\bibnamefont {Chen}}, \bibinfo
  {author} {\bibfnamefont {Y.}~\bibnamefont {Wu}}, \bibinfo {author}
  {\bibfnamefont {T.}~\bibnamefont {Wu}}, \bibinfo {author} {\bibfnamefont
  {L.}~\bibnamefont {Li}}, \bibinfo {author} {\bibfnamefont {N.}~\bibnamefont
  {Jiang}},\ and\ \bibinfo {author} {\bibfnamefont {Y.}~\bibnamefont {Wu}},\
  }\bibfield  {title} {\bibinfo {title} {Thermal anisotropy modulation as a
  major source of third harmonic magnetoresistance in ni thin films},\ }\href
  {https://doi.org/10.1103/ff9x-29xc} {\bibfield  {journal} {\bibinfo
  {journal} {Phys. Rev. B}\ }\textbf {\bibinfo {volume} {113}},\ \bibinfo
  {pages} {224413} (\bibinfo {year} {2026})}\BibitemShut {NoStop}%
\bibitem [{\citenamefont {Fedel}\ \emph {et~al.}(2025)\citenamefont {Fedel},
  \citenamefont {Villa}, \citenamefont {Damerio}, \citenamefont {Demiroglu},
  \citenamefont {Deger}, \citenamefont {Gazquez},\ and\ \citenamefont
  {Avci}}]{fedel2025AFM}%
  \BibitemOpen
  \bibfield  {author} {\bibinfo {author} {\bibfnamefont {S.}~\bibnamefont
  {Fedel}}, \bibinfo {author} {\bibfnamefont {M.}~\bibnamefont {Villa}},
  \bibinfo {author} {\bibfnamefont {S.}~\bibnamefont {Damerio}}, \bibinfo
  {author} {\bibfnamefont {E.}~\bibnamefont {Demiroglu}}, \bibinfo {author}
  {\bibfnamefont {C.}~\bibnamefont {Deger}}, \bibinfo {author} {\bibfnamefont
  {J.}~\bibnamefont {Gazquez}},\ and\ \bibinfo {author} {\bibfnamefont {C.~O.}\
  \bibnamefont {Avci}},\ }\bibfield  {title} {\bibinfo {title} {Evidence of
  long-range dzyaloshinskii--moriya interaction at ferrimagnetic
  insulator/nonmagnetic metal interfaces},\ }\href@noop {} {\bibfield
  {journal} {\bibinfo  {journal} {Advanced Functional Materials}\ ,\ \bibinfo
  {pages} {2418653}} (\bibinfo {year} {2025})}\BibitemShut {NoStop}%
\bibitem [{\citenamefont {L{\"o}w}\ \emph {et~al.}(2013)\citenamefont
  {L{\"o}w}, \citenamefont {Zvyagin}, \citenamefont {Ozerov}, \citenamefont
  {Schaufuss}, \citenamefont {Kataev}, \citenamefont {Wolf},\ and\
  \citenamefont {L{\"u}thi}}]{Low2013}%
  \BibitemOpen
  \bibfield  {author} {\bibinfo {author} {\bibfnamefont {U.}~\bibnamefont
  {L{\"o}w}}, \bibinfo {author} {\bibfnamefont {S.}~\bibnamefont {Zvyagin}},
  \bibinfo {author} {\bibfnamefont {M.}~\bibnamefont {Ozerov}}, \bibinfo
  {author} {\bibfnamefont {U.}~\bibnamefont {Schaufuss}}, \bibinfo {author}
  {\bibfnamefont {V.}~\bibnamefont {Kataev}}, \bibinfo {author} {\bibfnamefont
  {B.}~\bibnamefont {Wolf}},\ and\ \bibinfo {author} {\bibfnamefont
  {B.}~\bibnamefont {L{\"u}thi}},\ }\bibfield  {title} {\bibinfo {title}
  {{Magnetization, magnetic susceptibility and ESR in Tb3Ga5O12}},\ }\href@noop
  {} {\bibfield  {journal} {\bibinfo  {journal} {The European Physical Journal
  B}\ }\textbf {\bibinfo {volume} {86}},\ \bibinfo {pages} {1} (\bibinfo {year}
  {2013})}\BibitemShut {NoStop}%
\bibitem [{\citenamefont {Malozemoff}\ and\ \citenamefont
  {Slonczewski}(2013)}]{malozemoff2013magnetic}%
  \BibitemOpen
  \bibfield  {author} {\bibinfo {author} {\bibfnamefont {A.~P.}\ \bibnamefont
  {Malozemoff}}\ and\ \bibinfo {author} {\bibfnamefont {J.~C.}\ \bibnamefont
  {Slonczewski}},\ }\href@noop {} {\emph {\bibinfo {title} {Magnetic domain
  walls in bubble materials: advances in materials and device research}}},\
  Vol.~\bibinfo {volume} {1}\ (\bibinfo  {publisher} {Academic press},\
  \bibinfo {year} {2013})\BibitemShut {NoStop}%
\bibitem [{\citenamefont {Liu}\ \emph {et~al.}(2011)\citenamefont {Liu},
  \citenamefont {Moriyama}, \citenamefont {Ralph},\ and\ \citenamefont
  {Buhrman}}]{liu2011spin}%
  \BibitemOpen
  \bibfield  {author} {\bibinfo {author} {\bibfnamefont {L.}~\bibnamefont
  {Liu}}, \bibinfo {author} {\bibfnamefont {T.}~\bibnamefont {Moriyama}},
  \bibinfo {author} {\bibfnamefont {D.}~\bibnamefont {Ralph}},\ and\ \bibinfo
  {author} {\bibfnamefont {R.}~\bibnamefont {Buhrman}},\ }\bibfield  {title}
  {\bibinfo {title} {Spin-torque ferromagnetic resonance induced by the spin
  hall effect},\ }\href@noop {} {\bibfield  {journal} {\bibinfo  {journal}
  {Physical review letters}\ }\textbf {\bibinfo {volume} {106}},\ \bibinfo
  {pages} {036601} (\bibinfo {year} {2011})}\BibitemShut {NoStop}%
\end{thebibliography}%

\end{document}


\title{Supplemental Material: Giant nonlinear Hall effect in a Pt/ferrimagnetic insulator bilayer under Zeeman--exchange frustration}

\author{Takayuki~Shiino}
\email{tshiino@icmab.es}
\affiliation{Institut de Ciència de Materials de Barcelona (ICMAB-CSIC), Carrer dels Til.lers, 08193 Cerdanyola del Vallès, Spain}

\author{Matteo~Fettizio}
\affiliation{Institut de Ciència de Materials de Barcelona (ICMAB-CSIC), Carrer dels Til.lers, 08193 Cerdanyola del Vallès, Spain}

\author{Weronika~Janus}
\affiliation{Institut de Ciència de Materials de Barcelona (ICMAB-CSIC), Carrer dels Til.lers, 08193 Cerdanyola del Vallès, Spain}

\author{Can~Onur~Avci}
\email{cavci@icmab.es}
\affiliation{Institut de Ciència de Materials de Barcelona (ICMAB-CSIC), Carrer dels Til.lers, 08193 Cerdanyola del Vallès, Spain}

\date{\today}

\maketitle
\tableofcontents


\clearpage
\section{Experimental conditions}

\subsection{Sputtering conditions}

The heterostructure investigated in this study was prepared using the same growth procedure as that reported in our previous work \cite{Shiino2025tunable}. A nominally 23-nm-thick Al-substituted terbium iron garnet (Al:TbIG) layer was deposited on a Gd$_3$Ga$_5$O$_{12}$(111) [GGG(111)] substrate at 800~$^\circ$C by simultaneous radio-frequency (RF) magnetron sputtering from stoichiometric TbIG and Al$_2$O$_3$ targets, followed immediately by the deposition of a 2-nm-thick TbIG termination layer at the same temperature without breaking vacuum. Prior to deposition, the substrate was plasma-cleaned at room temperature using an excitation power of 30~W under an Ar pressure of 3~mTorr. During the co-sputtering process, the RF powers applied to the TbIG and Al$_2$O$_3$ targets were fixed at 150 and 87~W, respectively. The deposition was performed at a total working pressure of 3~mTorr using an Ar/O$_2$ gas-flow ratio of 30:2. The nominal thickness of the co-sputtered layer was determined from the independently calibrated deposition rate of TbIG, with a deposition time of approximately 1~h. After cooling the sample to room temperature, a 4-nm-thick Pt layer was deposited by direct-current magnetron sputtering at a power of 50~W under an Ar pressure of 3~mTorr, without exposing the sample to air. The resulting layer structure was Pt(4~nm)/TbIG(2~nm)/Al:TbIG(23~nm)//GGG(111), corresponding to a total garnet thickness of approximately 25~nm. The heterostructure was patterned into Hall-bar devices by optical photolithography and inductively coupled plasma reactive-ion etching.
The present Al:TbIG sample ($P_{\mathrm{Al_2O_3}}=87$~W ) exhibits perpendicular magnetic anisotropy at room temperature. The lattice constant of the present Al:TbIG layer was estimated to be $a_{\mathrm{Al:TbIG}}=12.424$~\AA, which is close to that of the GGG substrate ($a_{\mathrm{GGG}}=12.376$~\AA) \cite{Shiino2025tunable}. For comparison, the lattice constant of our undoped TbIG film grown on GGG(111) was $a_{\mathrm{TbIG}}=12.756$~\AA.

\subsection{Electrical measurement conditions}

\begin{figure*}[h]
\includegraphics[width=0.4\linewidth]{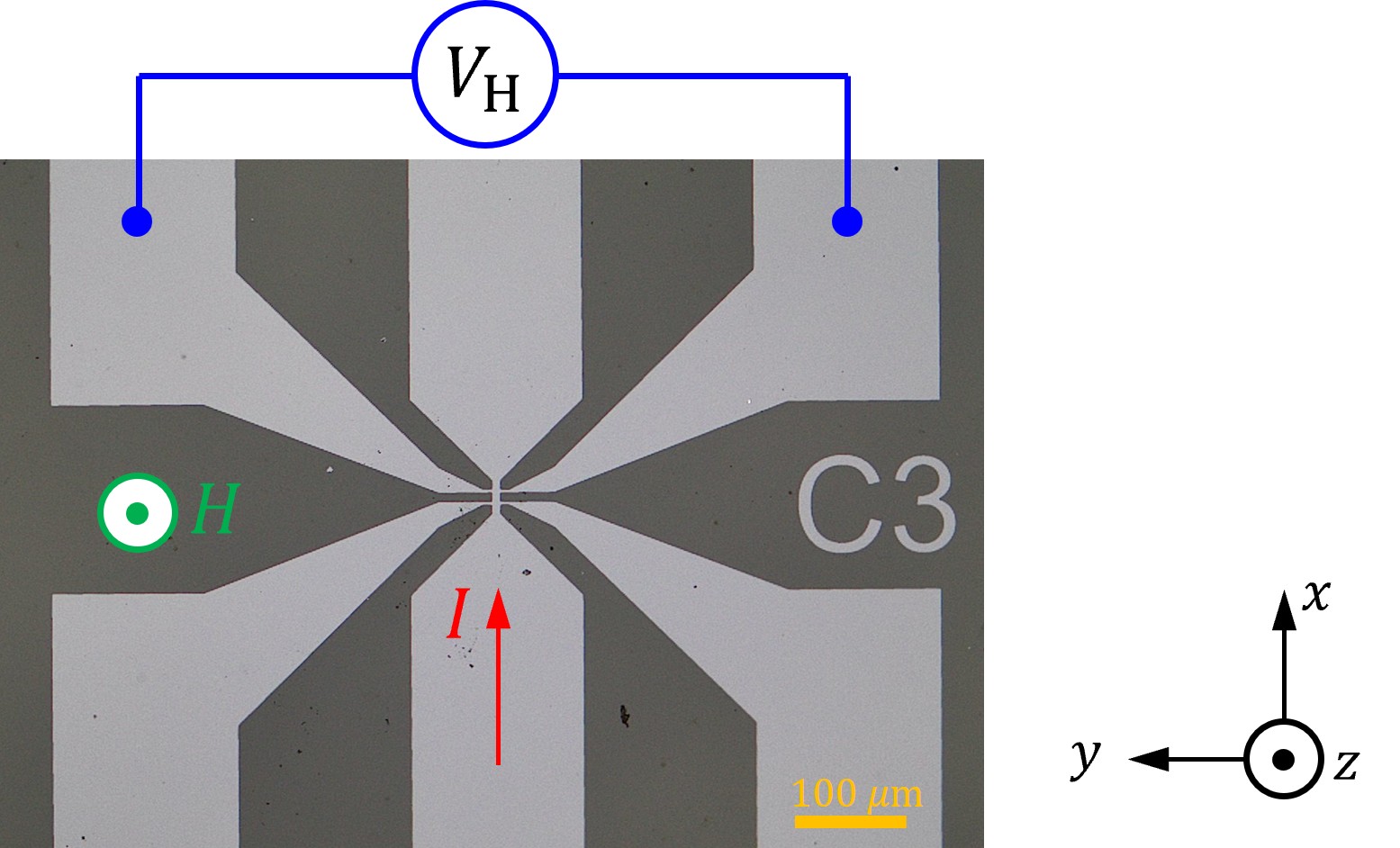}
\caption{Optical microscope image of the Hall-bar device and the electrical measurement geometry. The ac current $I$ was applied along the $x$ direction, the transverse Hall voltage $V_{\mathrm{H}}$ was measured along the $y$ direction, and the external magnetic field $H$ was applied along the $z$ direction, perpendicular to the film plane. The scale bar corresponds to 100~$\mu$m.}
\label{FIG_S_Hall_bar}
\end{figure*}

Electrical transport measurements were performed using an ac harmonic lock-in technique. Figure~\ref{FIG_S_Hall_bar} shows an optical microscope image of the Hall-bar device and the measurement geometry. A sinusoidal current $I(t)=I_0\sin(\omega t)$ was applied along the $x$ direction of the Hall bar, and the transverse Hall voltage $V_{\mathrm{H}}$ was measured along the $y$ direction. The external magnetic field $H$ was applied along the $z$ direction, perpendicular to the film plane. The transverse voltage was detected at integer multiples $n\omega$ of the excitation frequency. Both the in-phase (real) and quadrature (imaginary) components of the harmonic Hall voltages were recorded using MFLI lock-in amplifiers (Zurich Instruments).
Unless otherwise stated, the excitation frequency was $f=\omega/(2\pi)=337$~Hz. Because the garnet layers are electrically insulating, the applied current was assumed to flow entirely through the 4-nm-thick Pt layer. The current-density amplitude was calculated as $j_0=I_0/(w t_{\mathrm{Pt}})$, where $w=7.5~\mu\mathrm{m}$ is the Hall-bar channel width and $t_{\mathrm{Pt}}=4$~nm is the Pt thickness. Temperature-dependent measurements were conducted using the same custom-built heating and cooling setup as in our previous work \cite{Shiino2025tunable}. The linear ordinary Hall contribution was determined from the high-field region and subtracted from the measured first-harmonic Hall-resistance curves.

\clearpage
\section{Supplemental experimental data and analyses}

\subsection{Imaginary components of the harmonic Hall response}
\label{sec:SI_imaginary}

\begin{figure*}[h]
\includegraphics[width=0.8\linewidth]{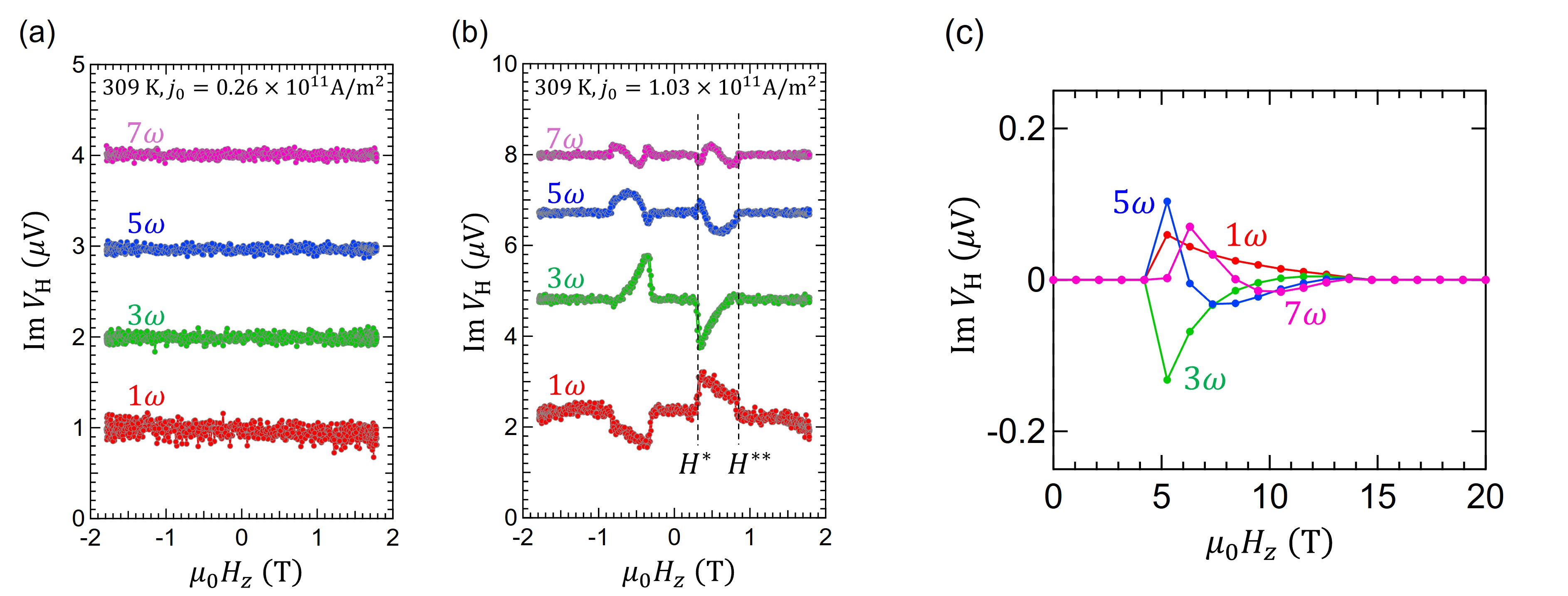}
\caption{
Imaginary components of the harmonic Hall response.
(a)--(b) Imaginary parts of the odd harmonic Hall voltages, \(V_H^{n\omega}\) with \(n=1,3,5,7\), measured under the same conditions as in Figs.~2 (a)--(b) of the main text: 
(a) \(T=309\) K, \(j_0=0.26\times10^{11}\) A/m\(^2\); and
(b) \(T=309\) K, \(j_0=1.03\times10^{11}\) A/m\(^2\). 
Vertical dashed lines indicate \(H^\ast\) and \(H^{\ast\ast}\). 
The curves are vertically offset for clarity.
(c) Simulated imaginary components under Joule-heating-induced parametric modulation. Simulation details are discussed in Section \ref{subsec:numerical_calc}.
Finite imaginary components appear in the switching regime, reflecting delayed magnetization dynamics.
}
\label{FIG:S_imaginary}
\end{figure*}

Figures~\ref{FIG:S_imaginary}(a)--(b) show the imaginary components of the odd harmonic Hall voltages measured under the same conditions as in Figs.~2 (a)--(b) of the main text. 
In the low-current regime [Fig.~\ref{FIG:S_imaginary}(a)], the imaginary components are essentially featureless. 
When large higher-harmonic signals appear in the in-phase response, clear imaginary components also emerge in the same nonlinear field region between \(H^\ast\) and \(H^{\ast\ast}\) [Fig.~\ref{FIG:S_imaginary}(b)], or above \(H^\ast\) when \(H^{\ast\ast}\) is outside the measurement window. 
The observed large imaginary-part signals may reflect slow dynamics associated with the switching behavior in the real sample, including domain nucleation, domain-wall propagation, and pinning effects. 

The spin-chain simulation with Joule-heating-induced parametric modulation also yields finite imaginary components in the switching regime [Fig.~\ref{FIG:S_imaginary}(e)]. 
In the simulation, the delay originates from the intrinsic damping in the Landau--Lifshitz--Gilbert dynamics and remains small because the response is nearly quasistatic. 


\clearpage
\subsection{Extent of the higher-order harmonic signals}

\begin{figure*}[h]
\includegraphics[width=0.5\textwidth]{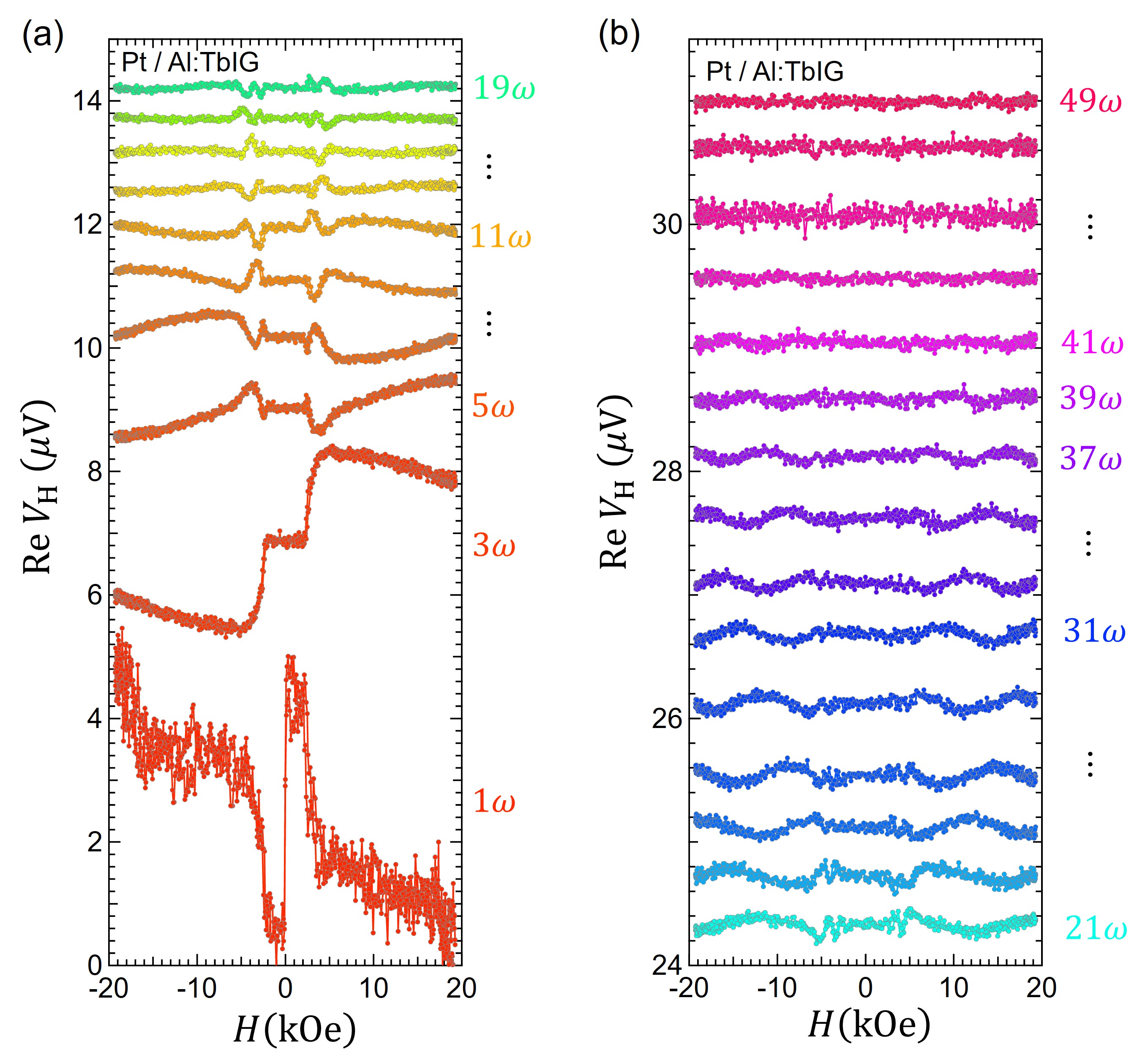}
\caption{
(a)--(b) Higher-order harmonic Hall measurement results for the Pt/Al:TbIG (25 nm) system at room temperature. 
The applied current density (amplitude) is $j_0 = 1.8 \times 10^{11}$ A/m$^{2}$.
(a) Odd-order harmonic signals from $1\omega$ to $19\omega$. 
(b) Odd-order harmonic signals from $21\omega$ to $49\omega$. 
All Hall voltage curves are vertically offset. 
}
\label{FIG_S_higher-order-harmonics_AlTbIG}
\end{figure*}

Figure \ref{FIG_S_higher-order-harmonics_AlTbIG}(a) and (b) show the odd-order harmonic signals of the main sample Pt/Al:TbIG (25 nm) up to $49\omega$ measured at room temperature. 
We observe clear nonlinear harmonic signals above the transition field of 2.5 kOe. 
Remarkably, the nonlinear signal is detectable up to around $37\omega$ harmonics, indicating the very large nonlinearity of the transport signal.

\begin{figure*}[h]
\includegraphics[width=0.3\textwidth]{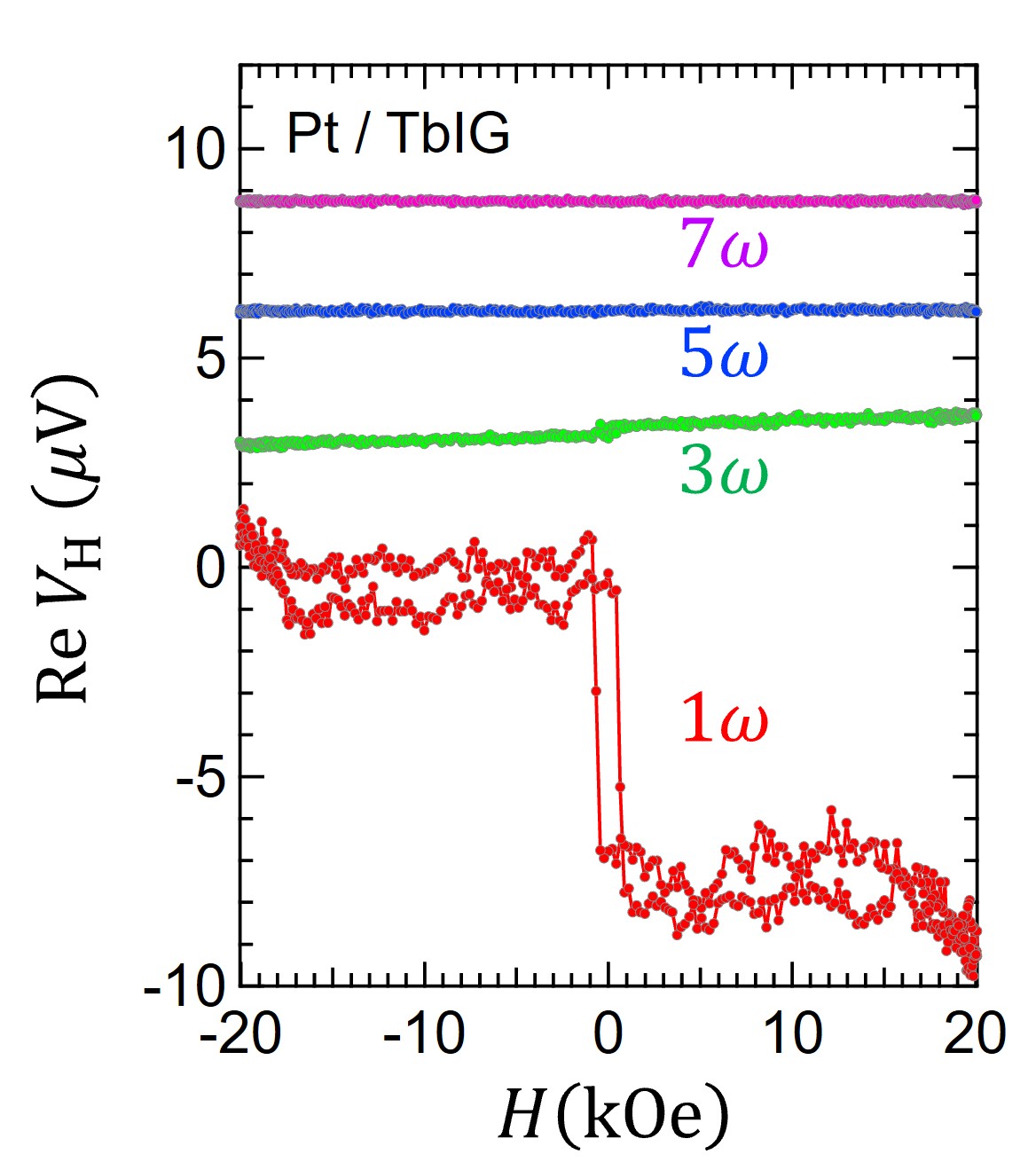}
\caption{
Odd-number harmonic signals of the Pt/TbIG (25 nm) reference system at room temperature.
The applied current density (amplitude) is $j_0 = 1.8 \times 10^{11}$ A/m$^{2}$.
The ordinary Hall effect component is subtracted in the $1\omega$ data.
All Hall voltage curves are vertically offset. 
}
\label{FIG_S_TbIG}
\end{figure*}

For comparison, in Fig. \ref{FIG_S_TbIG}, we show the odd-order harmonic Hall measurement result of the reference sample Pt/TbIG(25 nm) measured at room temperature, whose magnetic compensation temperature (190 K) is far from the measurement temperature. 
Only a  very small nonlinear $3\omega$ signal (at the switching field) is observed. 
There is no higher-order nonlinear harmonic signal. 

\clearpage
\subsection{Absence of even-number harmonic signals}

\begin{figure*}[h]
\includegraphics[width=0.3\textwidth]{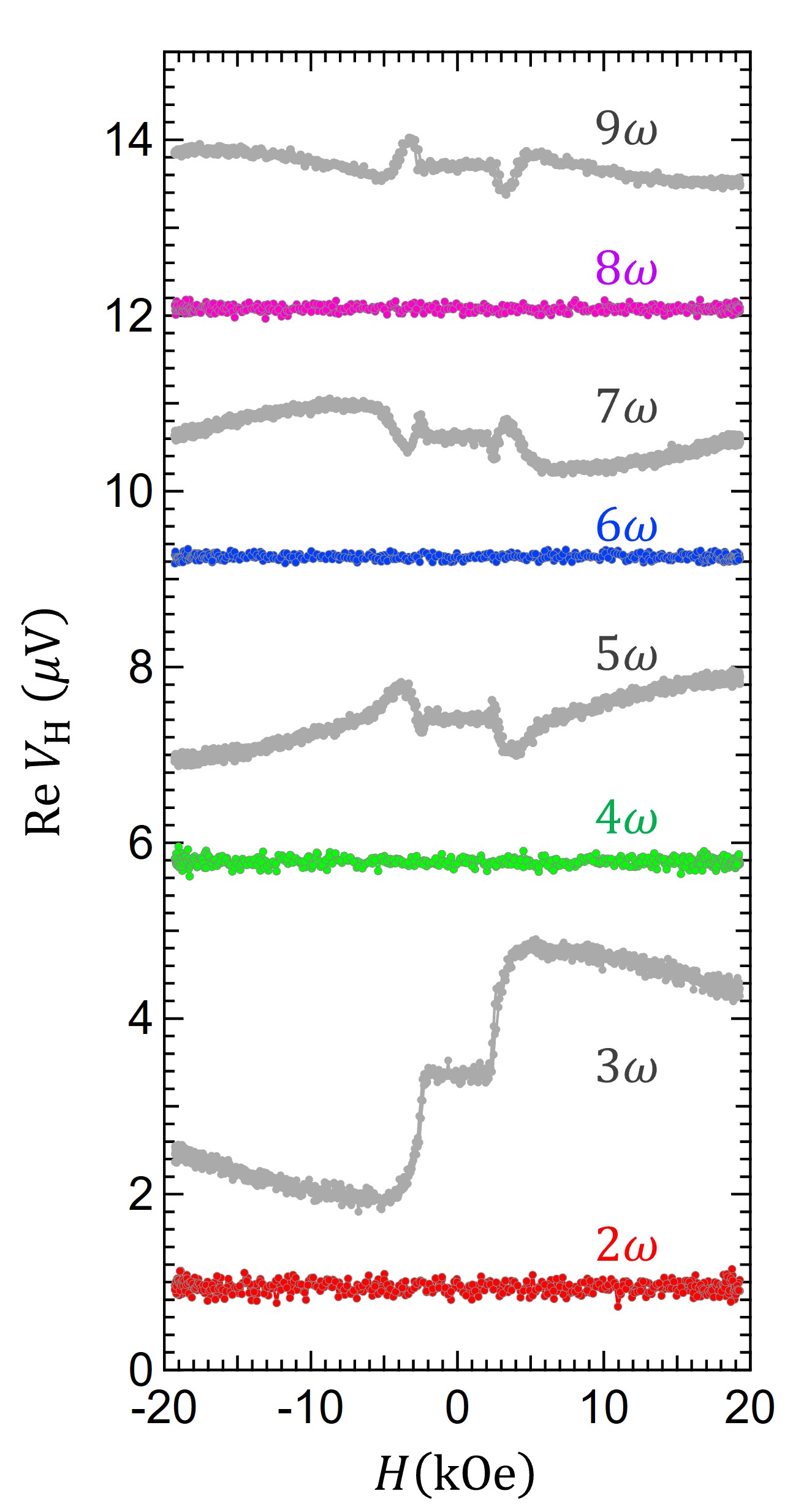}
\caption{
Even-order ($2\omega$, $4\omega$, $6\omega$ and $8\omega$) harmonic Hall measurement results for the Pt/Al:TbIG (25 nm) system at room temperature.
The applied current density (amplitude) is $j_0 = 1.8 \times 10^{11}$ A/m$^{2}$.
For comparison, the odd-order ($3\omega$, $5\omega$, $7\omega$ and $9\omega$) harmonic Hall curves are also plotted.
All Hall voltage curves are vertically offset. 
}
\label{FIG_S_even_harmonics}
\end{figure*}

Figure \ref{FIG_S_even_harmonics} shows the even-order harmonic Hall measurement results for the Pt/Al (25 nm) system at room temperature and a current density of $j_0 = 1.8 \times 10^{11}$ A/m$^{2}$, under the same conditions as those used for the data shown in Fig. \ref{FIG_S_higher-order-harmonics_AlTbIG}.
We do not observe any clear nonlinear signals at the transition (2.5 kOe) in the even-number harmonics. 

\clearpage
\subsection{Supplemental data collected for the phase diagram}

\begin{figure*}[h]
\includegraphics[width=0.7\textwidth]{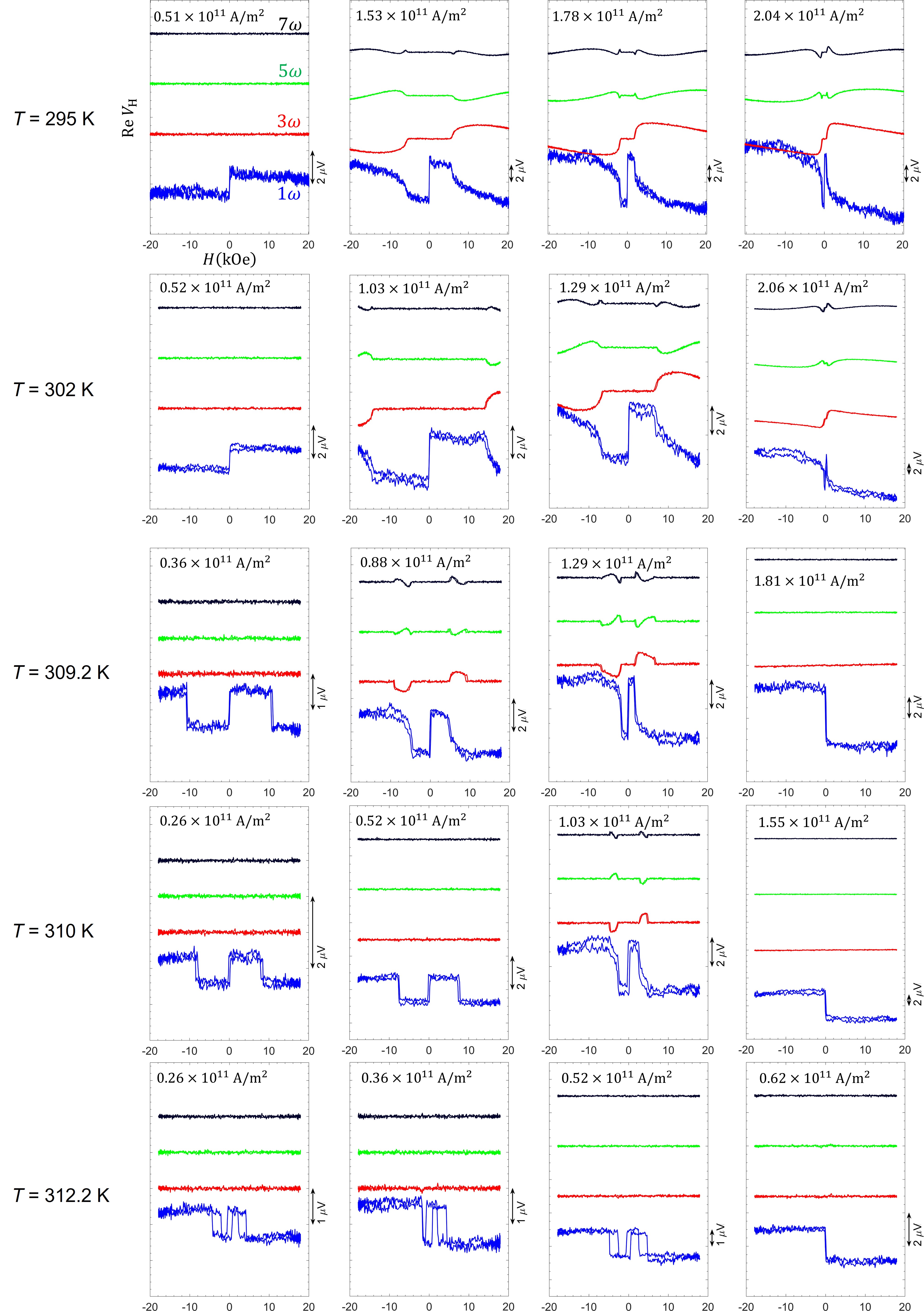}
\caption{
Selected data of the real part of AHE measured at different temperatures: (from top rows to bottom rows) 295 K, 302 K, 306.4 K, 309.2 K, 309.5 K, and 312.2 K. 
In each panel, $1\omega$ (blue), $3\omega$ (red), $5\omega$ (green), and $7\omega$ (black) harmonic data are presented. 
Applied current density is displayed in each panel: small current density (left) to higher current density (right). 
}
\label{FIG_S_AHE-DATA_selected}
\end{figure*}

We present selected AHE data used to plot $H^{*}$ and $H^{**}$ in Fig. 3 in the main text. 
Figure \ref{FIG_S_AHE-DATA_selected} shows the selected data of the real part of AHE measured at different temperatures: 295 K, 302 K, 306.4 K, 309.2 K, 309.5 K, and 312.2 K.

\clearpage
\subsection{Frequency dependence of the spin-flip transition field}

\begin{figure*}[h]
\includegraphics[width=0.8\textwidth]{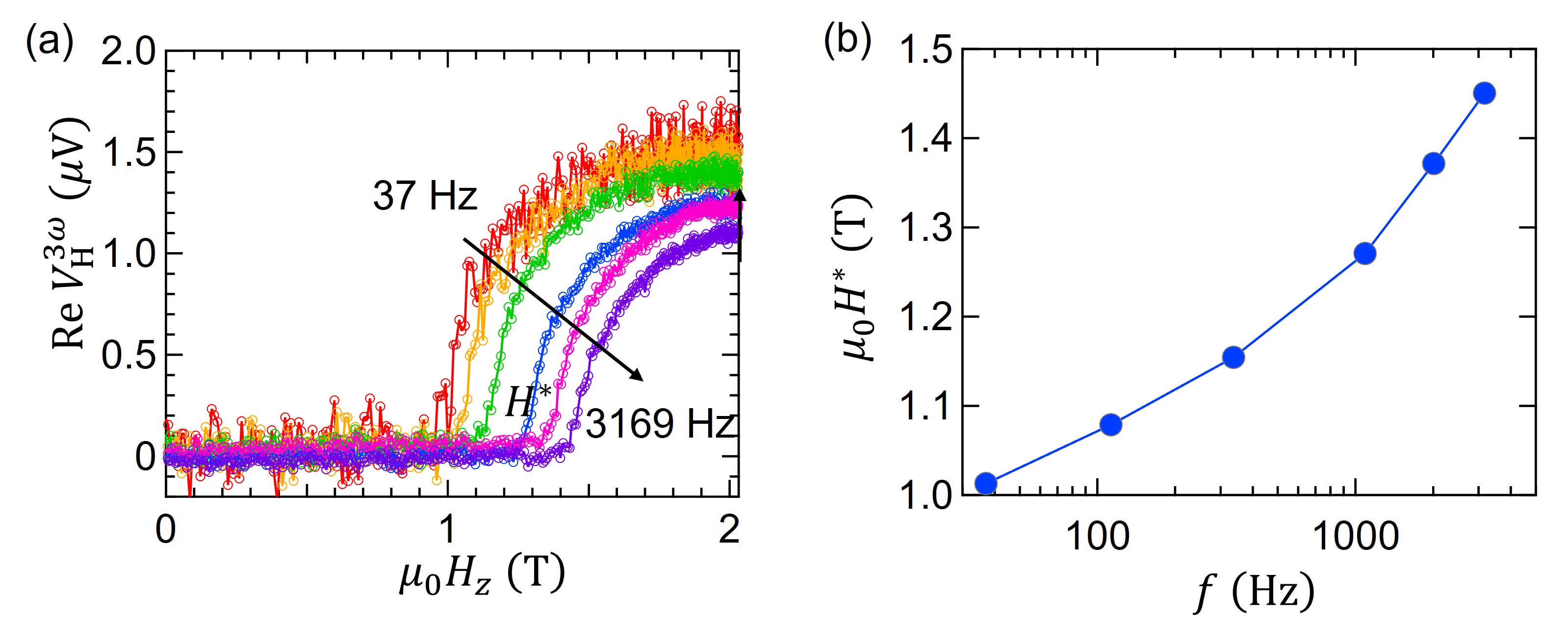}
\caption{(a) Field dependence of Re~$V^{3\omega}_{\rm H}$ for various ac frequencies $f$ measured at $T = 295$ K for $j_0 = 1.37\times 10^{11}$ A/m$^2$. 
The transition field $H^*$ is defined as the onset field where Re~$V^{3\omega}_{\rm H}$ starts to increase.
(b) Frequency dependence of $H^*$. 
The Hall voltage curves are vertically offset for clarity. 
}
\label{FIG_S:Nonequilibrium}
\end{figure*}

Figure~\ref{FIG_S:Nonequilibrium}(a) shows Re~$V^{3\omega}$ as a function of $H$ for different AC frequencies ($f$).
The transition field $H^{*}$ clearly shifts to higher values with increasing $f$, as also seen in Fig.~\ref{FIG_S:Nonequilibrium}(b).
The observed shift of $H^{*}$ likely reflects slow dynamics associated with magnetization switching, as also suggested by the imaginary component (i.e., the delayed response) of the nonlinear signals.

\clearpage
\subsection{Estimation of temperature increase by Joule heating effect}

\begin{figure*}[h]
\includegraphics[width=0.4\textwidth]{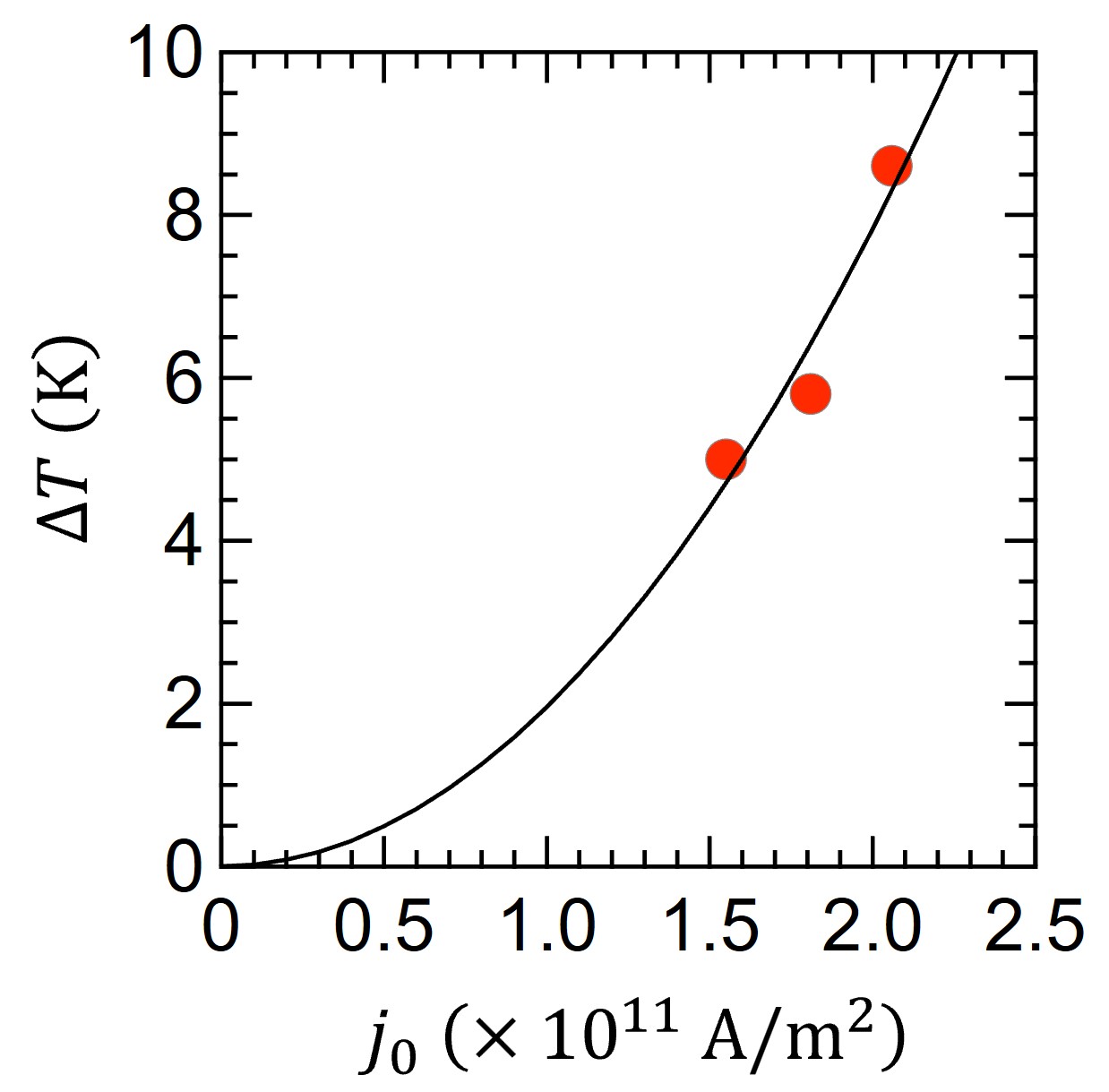}
\caption{
Estimated temperature increase ($\Delta T$) due to Joule heating, plotted as a function of the peak-to-peak amplitude of the ac current density ($j_0$). 
The solid curve represents a parabolic fit to the data, assuming $\Delta T \propto j_{0}^{2}$.
}
\label{FIG_S_Joule}
\end{figure*}

Here, we estimate the temperature increase due to Joule heating during transport measurements performed on the Hall-bar device.
From the sign change of the AHE curve near $H = 0$ under low current, we determined the effective compensation temperature to be $T_{\rm M} = 315$ K in the present system. 
Below $T_{\rm M}$, when a large current is applied and the local temperature increases due to Joule heating, the sign of the AHE near $H = 0$ reverses.
For example, as shown in Fig. \ref{FIG_S_AHE-DATA_selected}, at $T = 309.2$ K the system exhibits a positive AHE sign at low current ($0.36 \times 10^{11}$ A/m$^2$), whereas it shows a negative AHE sign at high current ($1.81 \times 10^{11}$ A/m$^2$).
This observation indicates that the system temperature increases by approximately ${\Delta}T \approx 6$ K—sufficient to cross $T_{\rm M} = 315$ K from the base temperature of 309.2 K—when a current density of $1.81 \times 10^{11}$ A/m$^2$ is applied.
This analysis allows us to estimate the current-density dependence of the Joule-heat-induced temperature rise.
Figure \ref{FIG_S_Joule} displays the estimated temperature increase (${\Delta}T$) as a function of the peak-to-peak amplitude of the ac current density ($j_0$).
In our experiment, strong nonlinear behavior emerged in the range of $0.5 \lesssim j_{0} \lesssim 2 \times 10^{11}$ A/m$^{2}$, where a Joule-heating-induced temperature rise of approximately $0.5 \lesssim \Delta T \lesssim 8$ K is estimated.

\clearpage
\subsection{Measurements on Al:TbIG films with different thickness and surface termination}

\begin{figure*}[h]
\includegraphics[width=0.8\textwidth]{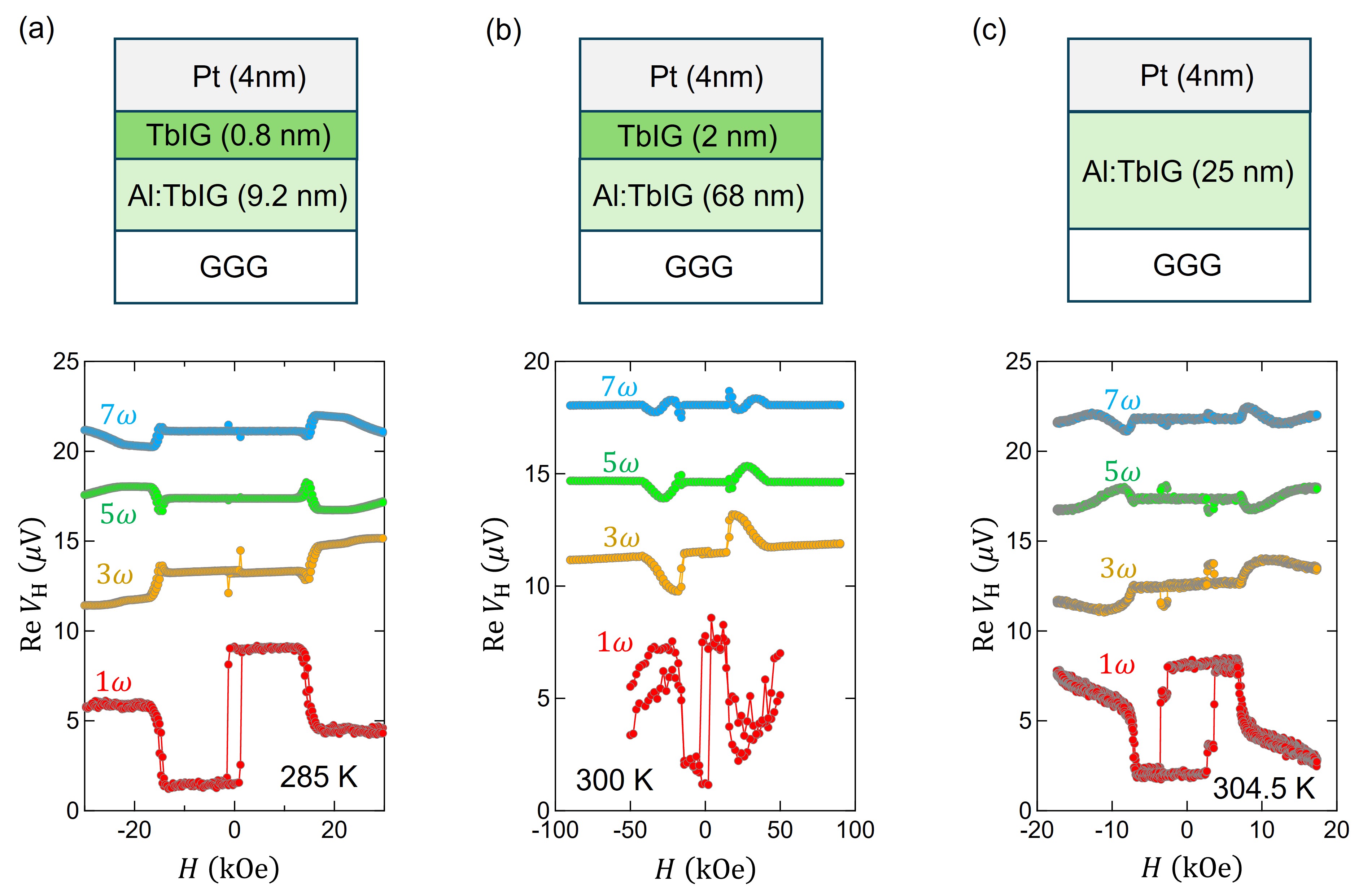}
\caption{
(a)–(c) Sample stack structures and representative harmonic Hall voltages ($1\omega$, $3\omega$, $5\omega$, and $7\omega$) of the reference samples: (a) 10-nm-thick Al:TbIG, (b) 70-nm-thick Al:TbIG, and (c) Al:TbIG without 2-nm TbIG termination. The Hall voltage data are vertically offset for clarity. Measurement temperatures are (a) 285 K, (b) 300 K, and (c) 304.5 K. Note that the top interfacial regions [(a) TbIG (0.8 nm) and (b) TbIG (2 nm), where only TbIG was sputtered] are also partially substituted by Al due to thermal annealing during sputtering.
}
\label{FIG_S_ref_samples}
\end{figure*}

Figures \ref{FIG_S_ref_samples}(a)–(c) show the nominal sample stack structures of the test samples and their representative harmonic Hall voltages ($1\omega$, $3\omega$, $5\omega$, and $7\omega$): (a) 10-nm-thick Al:TbIG, (b) 70-nm-thick Al:TbIG, and (c) a system without TbIG termination.
Note that these samples were prepared under conditions similar to those of the main 25-nm Al:TbIG sample, with slight tuning of the Al$_2$O$_3$ sputtering power during co-sputtering.
These test samples also exhibit similarly strong nonlinear response behavior at slightly different temperatures around room temperature.

The thinner 10-nm Al:TbIG sample (Fig. \ref{FIG_S_ref_samples}(a)) shows incomplete spin-flip behavior above the transition field ($H > H^{*}$) within our measurement range.
This may suggest that the spins in the top interfacial region are energetically stabilized in a canted configuration due to geometrical constraints on the formation of a compensation domain wall.

The thicker 70-nm Al:TbIG sample (Fig. \ref{FIG_S_ref_samples}(b)) also exhibits a field-induced spin-flip transition and distinct nonlinear Hall signals ($3\omega$, $5\omega$, and $7\omega$).
Note that the measured $1\omega$ data were noisy due to unidentified offset signals superimposed on the $1\omega$ voltage, presumably arising from poor electrical contact.

The sample without the 2-nm TbIG termination (Fig. \ref{FIG_S_ref_samples}(c)) also shows a field-induced transition and distinct nonlinear Hall signals ($3\omega$, $5\omega$, and $7\omega$).
This result may reflect an unintentionally developed compositional gradient in the sample. 
Such a gradient can arise even in stacks without deliberately introduced compositional variations and may have developed during the high-temperature growth of the Al:TbIG layer. 
However, a vertical compositional gradient is not the only possible origin of the Zeeman–exchange competition. Nanoscale lateral inhomogeneities could likewise generate competing magnetic regions and give rise to frustration-driven higher-harmonic signals.
Further investigation is required to determine the microscopic origin of the competing magnetic regions in this nominally homogeneous sample.


\clearpage
\section{Supplementary information for the numerical calculation}

\subsection{Numerical calculation condition of one-dimensional spin-chain model}
\label{subsec:numerical_calc}

\textbf{Model.} Under the assumption that the relevant variations of the physical parameters required to reproduce the experimental data occur only along the film thickness, we model the system as a one-dimensional chain of coupled spins. Each spin \( i \) is characterized by a saturation magnetization \( M_i \) and a unit vector \( \mathbf{m}_i \), and experiences an effective magnetic field \( \mathbf{H}_{\mathrm{eff},i} = \mathbf{H}_{\mathrm{EX},i} + \mathbf{H}_{\mathrm{AN},i} + \mathbf{H}_{\mathrm{Z},i} \). The nearest-neighbor exchange field is defined as \( \mathbf{H}_{\mathrm{EX},i} = \frac{J}{\mu_0 M_i}(\mathbf{m}_{i-1} + \mathbf{m}_{i+1}) \), where \( J \) is the volume density of the exchange energy. The uniaxial perpendicular anisotropy field is \( \mathbf{H}_{\mathrm{AN},i} = H_{\mathrm{AN}}(\mathbf{m}_i \cdot \mathbf{u}_Z)\mathbf{u}_Z = \frac{2K_{\mathrm{AN}}}{\mu_0 M_i} (\mathbf{m}_i \cdot \mathbf{u}_Z)\mathbf{u}_Z \), where \( K_{\mathrm{AN}} \) is the perpendicular anisotropy constant and \( \mathbf{u}_Z \) the out-of-plane unit vector. The Zeeman contribution reads \( \mathbf{H}_{\mathrm{Z},i}=\mathrm{sign}(M_i)\left(H_{\mathrm{EXT}}\mathbf{u}_Z+\delta_X\mathbf{u}_X+\delta_Y\mathbf{u}_Y\right) \), where the factor \(\mathrm{sign}(M_i)\) changes the sign of the Zeeman term depending on whether the local temperature is above or below the local magnetic compensation temperature of each spin, thereby accounting for the compensation-temperature gradient across the film thickness and reproducing the Zeeman--exchange frustration mechanism discussed in the main text. The coefficients \(\delta_{\rm X,Y}\) are randomly assigned to each spin from a normal distribution with standard deviation \(\sigma_{\rm H}=10^3\,\mathrm{A/m}\), representing thermal fluctuations of the effective field. As our focus is solely on their destabilizing effect, we adopt this simplified approach to reduce computational cost, rather than evaluating them via the fluctuation--dissipation theorem.

\textbf{Model parameters.} For clarity, all geometric, magnetic, transport, and numerical parameters used in the simulations are summarized in Table~\ref{tab:simulation_parameters}. The number of spins \(N\) is estimated from the effective number of Fe atomic planes, assuming a spacing \(a_{\rm Fe}=a_{\rm TbIG}/\sqrt[3]{40}\), where \(a_{\rm TbIG}\) is the TbIG lattice parameter and 40 the number of Fe atoms per unit cell. For \(t_{\rm TbIG}=25\,\mathrm{nm}\) and \(a_{\rm TbIG}=1.243\,\mathrm{nm}\), we obtain \(N=t_{\rm TbIG}/a_{\rm Fe}=69\). The top \(N_1\) spins are assigned a different compensation temperature, corresponding to the region with reduced Al content arising from the pure TbIG(2) insertion. Our previous EELS measurements (see Fig.~1 of the manuscript) indicate that diffusion broadens this region by approximately \(1\,\mathrm{nm}\), leading to \(N_1=3\,\mathrm{nm}/a_{\rm Fe}=8\). The anisotropy constant \(K_{\rm AN}\) is estimated using reported values of \(H_{\rm AN}\) and saturation magnetization \(M_{\rm S}\) for sputtered TbIG at room temperature (RT), namely \(H_{\rm AN}=2\,\mathrm{T}\) and \(M_{\rm S}=50\,\mathrm{kA/m}\) \cite{fedel2025AFM}. Since Al substitution reduces the lattice parameter and the anisotropy in TbIG is strain-induced, we rescale \(K_{\rm AN}\) to account for the reduced strain according to \(K_{\rm AN}(\mathrm{RT})=\frac{H_{\rm AN}\mu_0M_{\rm S}}{2}\left(\frac{a_{\mathrm{Al:TbIG}}-a_{\rm GGG}}{a_{\rm TbIG}-a_{\rm GGG}}\right)=7.5\,\mathrm{kJ/m^3}\). The temperature dependence is introduced through a linear coefficient \(dK/dT\), such that \(K(T)=K(\mathrm{RT})+\left(\frac{dK}{dT}\right)(T-\mathrm{RT})\), with \(dK/dT\simeq40\,\mathrm{J/m^3K}\), as reported for thin garnet films of comparable thickness \cite{song2024temperature}. The compensation temperatures of the top (\(T_{\rm cm,1}\)) and bottom (\(T_{\rm cm,2}\)) regions are determined from the global compensation temperature \(T_{\rm cm}=315\,\mathrm{K}\) and from the previously reported \(T_{\rm M}\)--\(P_{\rm AlOx}\) dependence (see Fig.~3(b) in Ref.~\cite{Shiino2025tunable}). Assuming linearity between AlOx sputtering power and Al content, the estimated \(18\%\) reduction in Al content in the top region corresponds to an effective sputtering power \(18\%\) smaller than the nominal \(87\,\mathrm{W}\). Using the extracted slope \(dT_{\rm cm}/dP_{\rm AlOx}\), this translates into a compensation-temperature difference \(\Delta T_{\rm cm}=33.5\,\mathrm{K}\). The temperature dependence of the layer magnetizations is described within a mean-field model under strong inter-sublattice exchange as \(M_{1(2)}(T)=M_{\rm Fe}(0)\left(1-\frac{T}{T_{\rm C}}\right)^{\beta}\left(\frac{T-T_{\rm cm,1(2)}}{T-T_{\rm W}}\right)\), where \(T_{\rm C}\approx400\,\mathrm{K}\) \cite{Shiino2025tunable}, \(\beta=0.33\), and \(T_{\rm W}=-7\,\mathrm{K}\) \cite{Low2013}. Accounting for the different magnetic volumes, the conditions \((N-N_1)M_2(T_{\rm cm})+N_1M_1(T_{\rm cm})=0\) and \(T_{\rm cm,1}=T_{\rm cm,2}-33.5\,\mathrm{K}\) are imposed. The exchange constant is obtained from the exchange stiffness \(A\) as \(J=2A/a_{\rm Fe}^2\). The room-temperature stiffness is estimated by scaling the YIG value (\(A_{\rm YIG}=4.15\,\mathrm{pJ/m}\) \cite{malozemoff2013magnetic}) according to the reduced Curie temperature: \(A(\mathrm{RT})=A_{\rm YIG}(\mathrm{RT})\frac{T_{\rm C}-\mathrm{RT}}{T_{\rm C,YIG}-\mathrm{RT}}\). The temperature dependence of \(J\) is taken as \(J(T)=J(\mathrm{RT})\left(\frac{T_{\rm C}-T}{T_{\rm C}-\mathrm{RT}}\right)^{2\beta}\). We note, however, that substantially better agreement between the experimental and simulated field dependence of the transitions is obtained by reducing the effective exchange constant \(J\) by a factor of six while leaving \(K_{\rm AN}\) unchanged. This phenomenological reduction partially accounts for the fact that, in the real system, magnetic transitions nucleate locally at defects characterized by lower values of \(J\) or \(K_{\rm AN}\) and proceed through thermally activated Arrhenius processes not included in the present model. Nevertheless, the agreement remains qualitative, as the simulated values of \(H^*\) and \(H^{**}\) are still approximately one order of magnitude larger than those measured experimentally.

\textbf{Spin--orbit torque and transport model.} Spin--orbit torque is introduced via the effective field \( \mathbf{H}_{\mathrm{SOT},i}=H_{\mathrm{DL},i}\mathbf{m}_i\times\mathbf{u}_y+H_{\mathrm{FL},i}\mathbf{u}_y \), with \(H_{\mathrm{DL},i}=\frac{\hbar}{2e}\frac{j_0\theta_{\rm SH}}{M_i a_{\rm Fe}}\) and \(H_{\mathrm{FL},i}=\beta H_{\mathrm{DL},i}\). Here, \(\theta_{\rm SH}\approx10\%\) \cite{liu2011spin} and \(\beta\sim0.1\). Since SOT is expected to dissipate within the first magnetic atomic layer of an insulator, \(\mathbf{H}_{\mathrm{SOT},i}\) is applied only to the top spin. The transverse spin Hall magnetoresistance is described by \(R_{\rm SMR}=R_{\rm A0}m_Z+R_{\rm P0}m_Xm_Y\), where \(R_{\rm A0}\) and \(R_{\rm P0}\) denote the anomalous Hall-like and planar Hall-like SMR coefficients, respectively. We use \(R_{\rm A0}=-0.48\,\mathrm{m\Omega}\), extracted experimentally, and \(R_{\rm P0}=0\), consistent with the multidomain character of the sample in zero in-plane magnetic field.

\textbf{Numerical implementation.} Joule heating is modeled through a coefficient \(dT/dj_0^2\), extracted from the current required to reach \(T_{\rm cm}\) at fixed external temperature. For example, at \(T=309.2\,\mathrm{K}\), \(j_0=1.81\times10^{11}\,\mathrm{A/m^2}\) increases the temperature by approximately \(6\,\mathrm{K}\), yielding \(dT/dj_0^2=3.7\times10^{-22}\,\mathrm{K/(A/m^2)^2}\). The Landau--Lifshitz equation is numerically solved for each spin as \(\frac{d\mathbf{m}_i}{dt}=-\gamma\,\mathbf{m}_i\times\mathbf{H}_{\mathrm{eff},i}-\gamma\,\alpha\,\mathbf{m}_i\times(\mathbf{m}_i\times\mathbf{H}_{\mathrm{eff},i})-\gamma\,\mathbf{m}_i\times\mathbf{H}_{\mathrm{SOT},i}\), with \(\alpha=0.01\). For AC simulations, a sinusoidal current \(j_0(t)=j_0\sin(\omega t)\) is applied, producing an SOT at frequency \(\omega\) and a temperature oscillation at \(2\omega\). The transverse voltage is computed from the top spin as \(V_{\rm H}(t)=j_0(t)AR_{\rm A0}m_{Z,1}(t)\). The \(n\)-th harmonic components are obtained via demodulation using in-phase and quadrature reference signals \(I_n(t)=\sin(n\omega t)\) and \(Q_n(t)=\cos(n\omega t)\), such that \(\mathrm{Re}(V_H^{n\omega})=\langle V_H(t)I_n(t)\rangle\) and \(\mathrm{Im}(V_H^{n\omega})=\langle V_H(t)Q_n(t)\rangle\).

\begin{table}[t]
\centering
\caption{Parameters used in the spin-chain simulations. Quantities labeled ``Experiment'' are obtained from measurements presented in this work; derived and estimated quantities are calculated as described in the text.}
\label{tab:simulation_parameters}
\begin{tabular}{llll}
\hline
Parameter & Symbol & Value & Origin \\
\hline
TbIG film thickness
& \(t_{\rm TbIG}\)
& \(25\,\mathrm{nm}\)
& Experiment \\

TbIG lattice parameter
& \(a_{\rm TbIG}\)
& \(1.243\,\mathrm{nm}\)
& Material parameter \\

Effective Fe-plane spacing
& \(a_{\rm Fe}\)
& \(0.364\,\mathrm{nm}\)
& \(a_{\rm TbIG}/\sqrt[3]{40}\) \\

Number of spins
& \(N\)
& 69
& \(t_{\rm TbIG}/a_{\rm Fe}\) \\

Thickness of the top region
& \(t_1\)
& \(3\,\mathrm{nm}\)
& EELS, Fig.~1 of the manuscript \\

Number of top-region spins
& \(N_1\)
& 8
& \(t_1/a_{\rm Fe}\) \\

Thermal-field standard deviation
& \(\sigma_{\rm H}\)
& \(10^3\,\mathrm{A/m}\)
& Assumed \\

Room-temperature anisotropy field
& \(H_{\rm AN}\)
& \(2\,\mathrm{T}\)
& Ref.~\cite{fedel2025AFM} \\

Room-temperature saturation magnetization
& \(M_{\rm S}\)
& \(50\,\mathrm{kA/m}\)
& Ref.~\cite{fedel2025AFM} \\

Room-temperature anisotropy constant
& \(K_{\rm AN}(\mathrm{RT})\)
& \(7.5\,\mathrm{kJ/m^3}\)
& Strain-rescaled estimate \\

Anisotropy temperature coefficient
& \(dK/dT\)
& \(40\,\mathrm{J\,m^{-3}\,K^{-1}}\)
& Ref.~\cite{song2024temperature} \\

Global compensation temperature
& \(T_{\rm cm}\)
& \(315\,\mathrm{K}\)
& Experiment \\

Nominal AlOx sputtering power
& \(P_{\rm AlOx}\)
& \(87\,\mathrm{W}\)
& Experiment \\

Top-region Al-content reduction
& \(\Delta c_{\rm Al}/c_{\rm Al}\)
& \(18\%\)
& Estimated from Fig.~1 \\

Compensation-temperature difference
& \(\Delta T_{\rm cm}\)
& \(33.5\,\mathrm{K}\)
& Estimated from Ref.~\cite{Shiino2025tunable} \\

Curie temperature
& \(T_{\rm C}\)
& \(400\,\mathrm{K}\)
& Ref.~\cite{Shiino2025tunable} \\

Magnetization critical exponent
& \(\beta\)
& 0.33
& Mean-field parametrization \\

Weiss temperature
& \(T_{\rm W}\)
& \(-7\,\mathrm{K}\)
& Ref.~\cite{Low2013} \\

YIG exchange stiffness at RT
& \(A_{\rm YIG}(\mathrm{RT})\)
& \(4.15\,\mathrm{pJ/m}\)
& Ref.~\cite{malozemoff2013magnetic} \\

Exchange-energy density
& \(J(T)\)
& \(J(\mathrm{RT})
\left(\frac{T_{\rm C}-T}
{T_{\rm C}-\mathrm{RT}}\right)^{2\beta}\)
& Derived from \(A\) \\

Effective exchange reduction factor
& \(J/J_{\rm estimated}\)
& \(1/6\)
& Phenomenological adjustment \\

Gilbert damping parameter
& \(\alpha\)
& 0.01
& Assumed \\

Spin Hall angle
& \(\theta_{\rm SH}\)
& 0.10
& Ref.~\cite{liu2011spin} \\

Field-like to damping-like SOT ratio
& \(\beta_{\rm SOT}\)
& 0.1
& Assumed \\

Anomalous Hall-like SMR coefficient
& \(R_{\rm A0}\)
& \(-0.48\,\mathrm{m\Omega}\)
& Experiment \\

Planar Hall-like SMR coefficient
& \(R_{\rm P0}\)
& 0
& Assumed from the multidomain state \\

Joule-heating coefficient
& \(dT/dj_0^2\)
& \(3.7\times10^{-22}\,
\mathrm{K/(A/m^2)^2}\)
& Experiment \\

Reference temperature for heating estimate
& \(T\)
& \(309.2\,\mathrm{K}\)
& Experiment \\

Reference current density
& \(j_0\)
& \(1.81\times10^{11}\,\mathrm{A/m^2}\)
& Experiment \\

Temperature increase at reference current
& \(\Delta T\)
& \(6\,\mathrm{K}\)
& Experiment \\
\hline
\end{tabular}
\label{tab1}
\end{table}

\clearpage
\subsection{Current and temperature dependent simulations}

\begin{figure*}[h]
\includegraphics[width=0.95\textwidth]{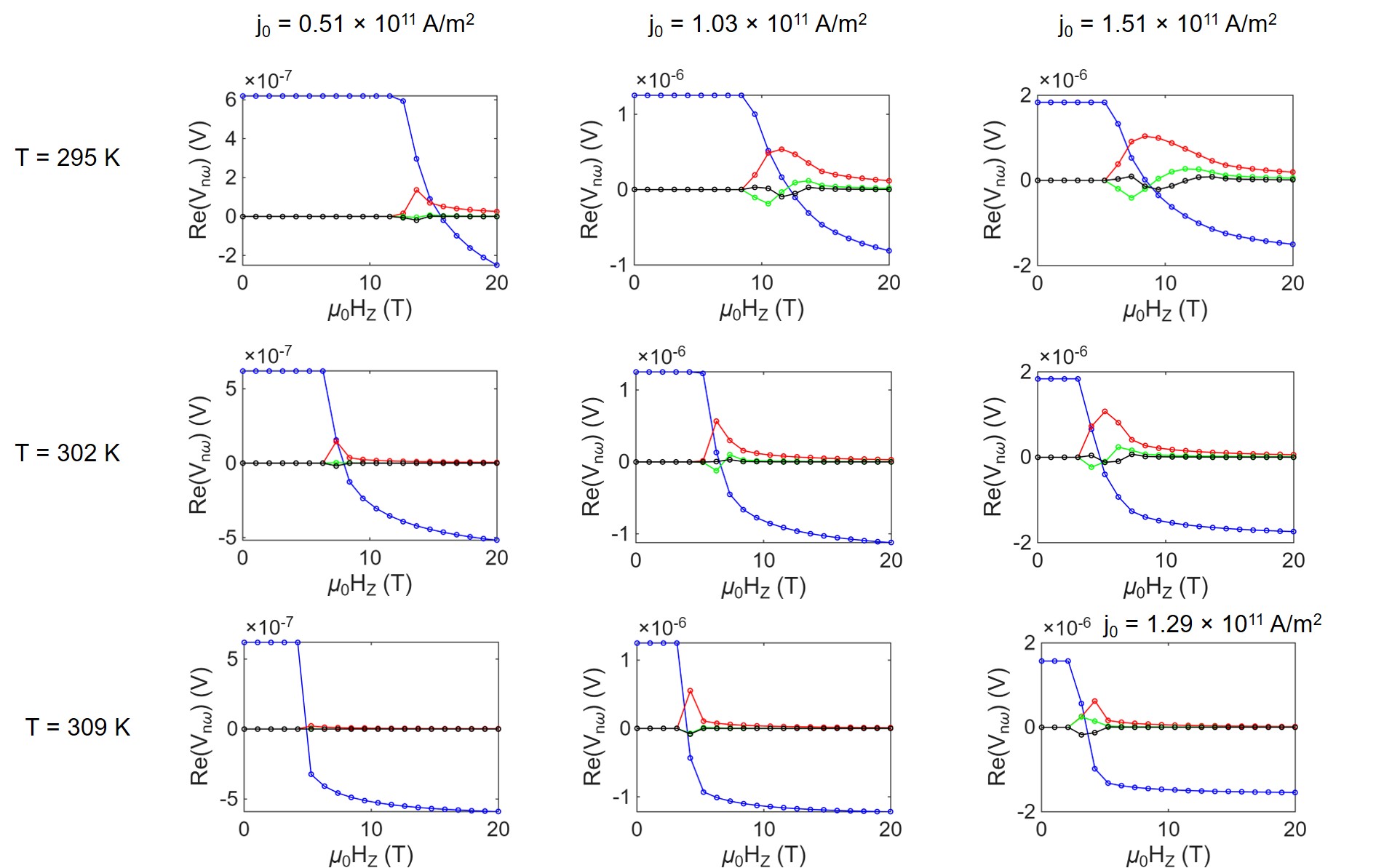}
\caption{real parts of the $f$ (blue curves), $3f$ (red curves), $5f$ (green curves), and $7f$ (black curves) harmonic components of the TSMR response for the various combinations of $T$ and $j_0$.}
\label{FIG_S_jDep_Tdep}
\end{figure*}

To further validate the model under different physical conditions, we performed AC simulations at $T = 295$, $302$, and $309$~K, using current densities $j_0 = 0.51\times10^{11}$, $1.03\times10^{11}$, and $1.51\times10^{11}$~A/m$^2$. At 309 K, however, $j_0 =1.51\times10^{11}$~A/m$^2$ brings the system too close to $T_{\rm M}$ and the simulation fails to converge toward a physical solution. For this reason, to show the effect of increasing $j_0$ at $T$ = 309 K, we set $j_0 =1.29\times10^{11}$~A/m$^2$ as the maximum value for $j_0$. Figure \ref{FIG_S_jDep_Tdep} displays the real parts of the $f$ (blue curves), $3f$ (red curves), $5f$ (green curves), and $7f$ (black curves) harmonic components of the SMR response for the various combinations of $T$ and $j_0$.

A qualitative agreement with the experimental observations is achieved. At a fixed temperature, increasing $j_0$ shifts the transition toward lower values of $H_{\rm Z}$ and broadens the transition window due to the enhanced average heating and larger oscillation amplitude associated with Joule heating. Conversely, for a fixed current density $j_0$, increasing the temperature reduces the transition field and narrows the transition window. This trend is attributed to the progressive reduction of the compensation domain wall as the system approaches the compensation temperature.

\clearpage
\subsection{Effect of spin--orbit torque}
\label{sec:SI_SOT}

\begin{figure}[h]
\includegraphics[width=0.8\linewidth]{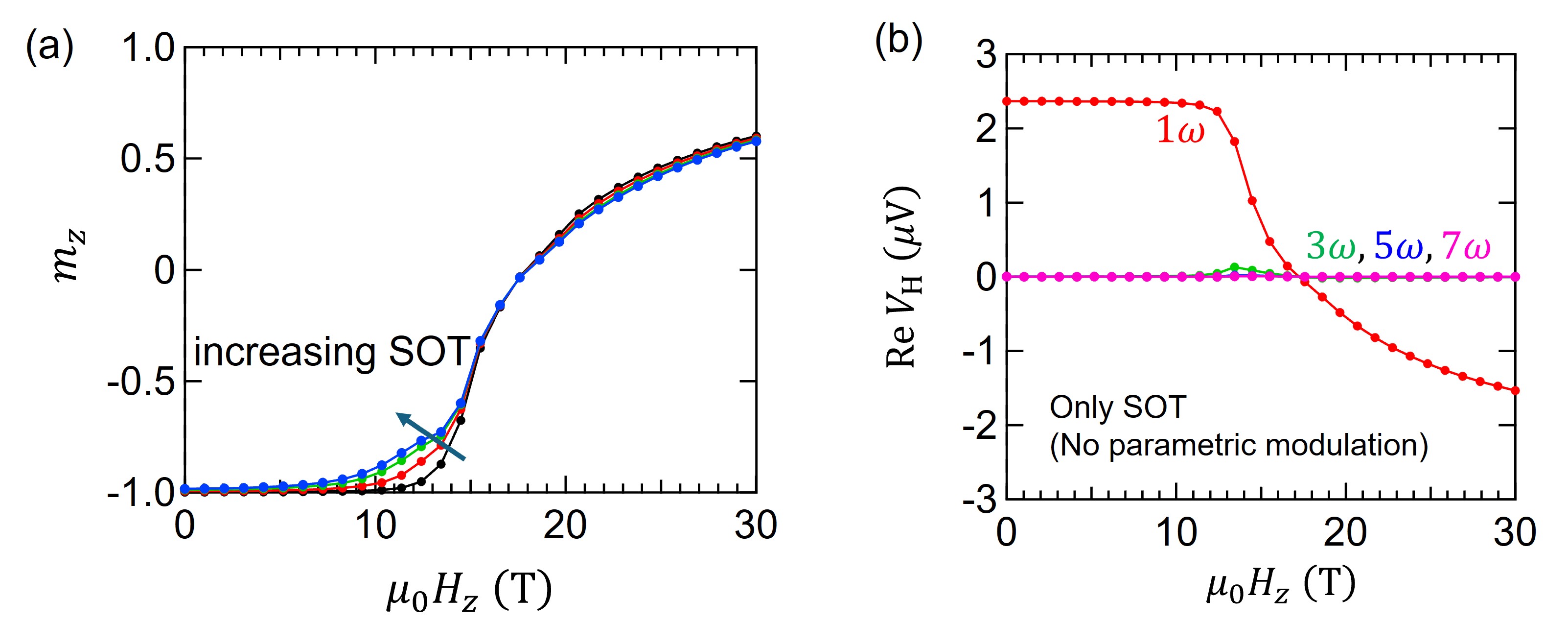}
\caption{
Effect of spin--orbit torque in the spin-chain simulation.
(a) Static out-of-plane magnetization \(m_z\) of the interfacial spin as a function of \(\mu_0H_z\) for increasing dc SOT. 
SOT slightly broadens the spin-reorientation region by assisting canting near the spin-flip transition.
(b) Simulated in-phase odd harmonic Hall voltages calculated with ac SOT only, without Joule-heating-induced parametric modulation. 
The higher harmonics remain negligible, showing that SOT alone cannot account for the giant nonlinear response observed experimentally.
}
\label{FIG:S_SOT}
\end{figure}

Close to the Zeeman--exchange instability, the effective magnetic field becomes sufficiently small that spin--orbit torque (SOT) may significantly influence the magnetization dynamics. To assess its contribution independently of thermal effects, we performed spin-chain simulations including SOT while neglecting Joule heating.
SOT was included as a current-induced torque acting on the top interfacial spin. 
As shown in Fig.~\ref{FIG:S_SOT}(a), a dc SOT slightly broadens the spin-reorientation region: near the spin-flip transition, the out-of-plane restoring field is weakened by Zeeman--exchange competition, so that SOT can assist canting of the interfacial magnetization.

We then calculated the harmonic response with ac SOT while suppressing Joule-heating-induced modulation of the magnetic parameters. 
The result is shown in Fig.~\ref{FIG:S_SOT}(b). 
The first-harmonic signal follows the field-driven spin reorientation, but the \(3\omega\), \(5\omega\), and \(7\omega\) components remain negligible. 
This contrasts with the experiment, where large higher-order odd harmonics appear over an extended field range. 
Therefore, SOT may weakly assist the destabilization of the collinear state, but it cannot by itself generate the giant nonlinear response; Joule-heating-induced parametric modulation is essential.

\clearpage

\bibliography{REFERENCE}